\documentclass{aa}

\usepackage{graphicx}
\usepackage{txfonts}
\usepackage{hyperref}
\usepackage{soul}
\usepackage{xcolor}
\usepackage{subcaption}
\usepackage{caption}
\usepackage[utf8]{inputenc}
\usepackage{natbib,twoopt}
\usepackage{siunitx}
\bibpunct{(}{)}{;}{a}{}{,}
\usepackage{silence}
\usepackage{ulem}

\begin{document} 

\WarningFilter{natbib}{Citation `Bommier2013' multiply defined}
\WarningFilter{natbib}{Citation `Asensio2008' multiply defined}
\WarningFilter{natbib}{Citation `Avrett1994' multiply defined}
\WarningFilter{natbib}{Citation `Balthasar2018' multiply defined}
\WarningFilter{natbib}{Citation `Bellot2004' multiply defined}
\WarningFilter{natbib}{ Citation `Benko2018' multiply defined}
\WarningFilter{natbib}{ Citation `Benko2021' multiply defined}
\WarningFilter{natbib}{Citation `Borrero2011' multiply defined}
\WarningFilter{natbib}{Citation `Collados1999' multiply defined}
\WarningFilter{natbib}{Citation `Collados2012' multiply defined}
\WarningFilter{natbib}{Citation `Collados2003' multiply defined}
\WarningFilter{natbib}{Citation `Denker2012' multiply defined}
\WarningFilter{natbib}{Citation `Manrique2020' multiply defined}
\WarningFilter{natbib}{Citation `Hale1908' multiply defined}
\WarningFilter{natbib}{Citation `Hofmann2012'  multiply defined}
\WarningFilter{natbib}{Citation `Joshi2014' multiply defined}
\WarningFilter{natbib}{Citation `Joshi2017' multiply defined}
\WarningFilter{natbib}{Citation `Joshi2016' multiply defined}
\WarningFilter{natbib}{Citation `Jurcak2011' multiply defined}
\WarningFilter{natbib}{Citation `Balthasar2013' multiply defined}
\WarningFilter{natbib}{Citation `Balthasar2008'  multiply defined}
\WarningFilter{natbib}{Citation `Mathew2003'  multiply defined}
\WarningFilter{natbib}{Citation `Merenda2011'  multiply defined}
\WarningFilter{natbib}{Citation `Moran2000' multiply defined}
\WarningFilter{natbib}{Citation `Orozco2005' multiply defined}
\WarningFilter{natbib}{Citation `Penn1995' multiply defined}
\WarningFilter{natbib}{Citation `Reiners2016' multiply defined}
\WarningFilter{natbib}{Citation `Ruedi1995' multiply defined}
\WarningFilter{natbib}{Citation `RuizCobo1992' multiply defined}
\WarningFilter{natbib}{Citation `SanchezCuberes2005' multiply defined}
\WarningFilter{natbib}{Citation `Schad2015' multiply defined}
\WarningFilter{natbib}{Citation `Schmidt2000' multiply defined}
\WarningFilter{natbib}{Citation `schmidt2012' multiply defined}
\WarningFilter{natbib}{Citation `Solanki2003' multiply defined}
\WarningFilter{natbib}{Citation `Tiwari2015' multiply defined}
\WarningFilter{natbib}{Citation `Tiwari2013' multiply defined}
\WarningFilter{natbib}{Citation `Verma2012' multiply defined}
\WarningFilter{natbib}{Citation `Westendorp1998' multiply defined}
\WarningFilter{natbib}{Citation `Westendorp2001' multiply defined}
\WarningFilter{natbib}{Citation `Westendorp2001b' multiply defined}
\WarningFilter{natbib}{Citation `Lagg2004' multiply defined}
\WarningFilter{natbib}{Citation `Xu2012' multiply defined}
\WarningFilter{natbib}{Citation `EST' multiply defined}
\WarningFilter{natbib}{Citation `Panos2018' multiply defined}
\WarningFilter{natbib}{Citation `macqueen1967' multiply defined}
\WarningFilter{natbib}{Citation `Pietarila2007' multiply defined}
\WarningFilter{natbib}{Citation `DKIST' multiply defined}
\WarningFilter{natbib}{Citation `Robustini2019' multiply defined}
\WarningFilter{natbib}{Citation `Schad2011' multiply defined}
\WarningFilter{natbib}{Citation `Socas2015' multiply defined}
\WarningFilter{natbib}{Citation `Solanki2003b' multiply defined}
\WarningFilter{natbib}{Citation `Sowmya2022' multiply defined}
\WarningFilter{natbib}{Citation `Viticchie2011' multiply defined}
\WarningFilter{natbib}{Citation `Yadav2019' multiply defined}
\WarningFilter{natbib}{Citation `Lagg2007' multiply defined}
\WarningFilter{natbib}{Citation `Kuckein2020' multiply defined}
\WarningFilter{natbib}{Citation `GonzalezManrique2016' multiply defined}
\WarningFilter{natbib}{Citation `GonzalezManrique2018' multiply defined}
\WarningFilter{natbib}{Citation `Felipe2023' multiply defined}
\WarningFilter{natbib}{Citation `delaCruzRodr2019' multiply defined}
\WarningFilter{natbib}{Citation `Carlsson2019' multiply defined}
\WarningFilter{natbib}{Citation `Bruzek1969' multiply defined}
\WarningFilter{natbib}{Citation `Bruzek1967' multiply defined}
\WarningFilter{natbib}{Citation `DiazBaso2019a' multiply defined}
\WarningFilter{natbib}{Citation `DiazBaso2019b' multiply defined}
\WarningFilter{natbib}{Citation `GonzalezManrique2024' multiply defined}
\WarningFilter{natbib}{Citation `Rast2021' multiply defined}
\WarningFilter{natbib}{Citation `Romano2023' multiply defined}
\WarningFilter{natbib}{Citation `sainz_dalda2019' multiply defined}
\WarningFilter{natbib}{Citation `Lindner2023' multiply defined}
\WarningFilter{natbib}{Citation `Grossmann-Doerth1994'' multiply defined}
\WarningFilter{natbib}{Citation `M4' multiply defined}
\WarningFilter{natbib}{Citation `Libbrecht2021' multiply defined}
\WarningFilter{natbib}{Citation `LitesSkumanich1990' multiply defined}
\WarningFilter{natbib}{Citation `Ruedi1992' multiply defined}
\WarningFilter{natbib}{Citation `T93-27' multiply defined}
\WarningFilter{hyperref}{Suppressing link with empty target on input line 178.}

  \title{The dependence of the magnetism of a near-limb sunspot on height}
    \author{
        M.\ Benko\inst{1},
        H.\ Balthasar\inst{2},
        P.\ G{\"o}m{\"o}ry\inst{1},
        C.\ Kuckein\inst{3,4,5},
        and S.J.\ Gonz{\'a}lez Manrique\inst{1, 6}
        }

    \authorrunning{Benko et al.}
   
    \institute{Astronomical Institute, Slovak Academy of Sciences, 
              Tatransk\'{a} Lomnica, Slovakia\\
              \email{mbenko@ta3.sk, gomory@ta3.sk}
        \and 
              Leibniz-Institute for Astrophysics Potsdam, Germany\\
        \and      
              Instituto de Astrof\'{i}sica de Canarias (IAC), V\'{i}a L\'{a}ctea s/n, 38205 La Laguna, Tenerife, Spain\\
        \and
              Departamento de Astrof\'{\i}sica, Universidad de La Laguna, 38205, La Laguna, Tenerife, Spain \\ 
        \and  
              Max-Planck-Institut f\"ur Sonnensystemforschung, Justus-von-Liebig-Weg 3, 37077 G\"ottingen, Germany \\
        \and  
              Institut für Sonnenphysik (KIS), Schöneckstr. 6, 79104 Freiburg, Germany\\
              }
    
  \date{Received November 13, 2023; accepted November 14, 2023}

\abstract
{The physical parameters of the sunspot are not fully understood, especially the height dependence of the magnetic field. So far, it is also an open question as to which heights the \ion{He}{i}\,10830\,\AA{} spectral line is formed at.}
{Our aim is to investigate the magnetic and dynamical properties in the atmosphere above a sunspot, from the photosphere to the chromosphere. We analyzed the photospheric and chromospheric magnetic field properties of a stable sunspot in AR 12553 on June 20, 2016 using spectropolarimetric observations obtained with the GREGOR Infrared Spectrograph (GRIS) at the 1.5-meter GREGOR telescope. }
{A spectral-line inversion technique was used to infer the magnetic field vector and Doppler velocities from the full Stokes profiles. In total, three spectral lines were inverted and the variation of the magnetic properties was qualified using the average values of the radial circles. The sunspot is located close to the solar limb, and thus this allows us to make a geometrical determination of the height of the spectral line \ion{He}{i}\,10830\,\AA.}
{We find the height of helium spectral line to be 970 km above the photospheric spectral lines directly from observation at a stable sunspot. The total magnetic field strength decreases with height over the sunspot; the rates are $-0.34\,\rm{G}\,\rm{km}^{-1}$ for the umbra and $-0.28\,\rm{G}\,\rm{km}^{-1}$ for the penumbra. The inclination increases with increasing height in the umbra, but decreases in the penumbra.  
In the umbra, the vertical component ($B_z$) decreases with height, while the horizontal component ($B_{hor}$) remains almost constant. In the penumbra this is reversed, as $B_z$ remains nearly constant over height, while $B_{hor}$ decreases. We also observe fast velocities with 30\,$\rm{km\,s}^{-1}$ in small chromospheric patches on the central side of the spot.}
{ The key parameters depending on height in the sunspot are the $B_{z}$ component of the magnetic field for the umbra and the $B_{hor}$ component of the magnetic field for the penumbra. The observation revealed supersonic downward velocities in and near the outer penumbra.}

\maketitle
\section{Introduction}
The magnetic field is the principal driver of physical processes in the solar atmosphere, with its strength and vector being the main aspect that characterizes active regions (ARs). The most dominant magnetic structures in ARs are sunspots \citep{Hale1908}.\newline
The photospheric appearance of a sunspot is characterized by a dark umbral core surrounded by a filamentary penumbra. Typically, the magnetic field strength and inclination of a sunspot are studied via spectral lines that have a high sensitivity (a large Landé factor, $g_{eff}$) to the presence of a magnetic field. The magnetic field of sunspots is typically analyzed by means of inversion codes. A large number of studies have investigated the global properties of the magnetic field in sunspots (for an overview see the reviews by \citet{Solanki2003} and \citet{Borrero2011}, and for detailed results see \citet{Westendorp1998, Moran2000, Bellot2004, SanchezCuberes2005, Orozco2005, Balthasar2008, Jurcak2011, Balthasar2013, Bommier2013, Socas2015, Tiwari2015, Joshi2017, Benko2018, Benko2021}, most of which focus on photospheric lines). The magnetic field of the umbra is stronger and more vertical than the penumbra. Sunspots are visible structures in both layers, the photosphere and the chromosphere \citep{Carlsson2019}. Investigations of the chromospheric magnetic field have also been performed and often focused on the He I infra-red triplet around 10830\,\AA{} \citep{Merenda2011, Schad2015, Xu2012, Joshi2016}.

\begin{figure}[hpt]
\includegraphics[width=\columnwidth, trim=0.8cm 0.35cm 1.5cm 1cm, clip=true]{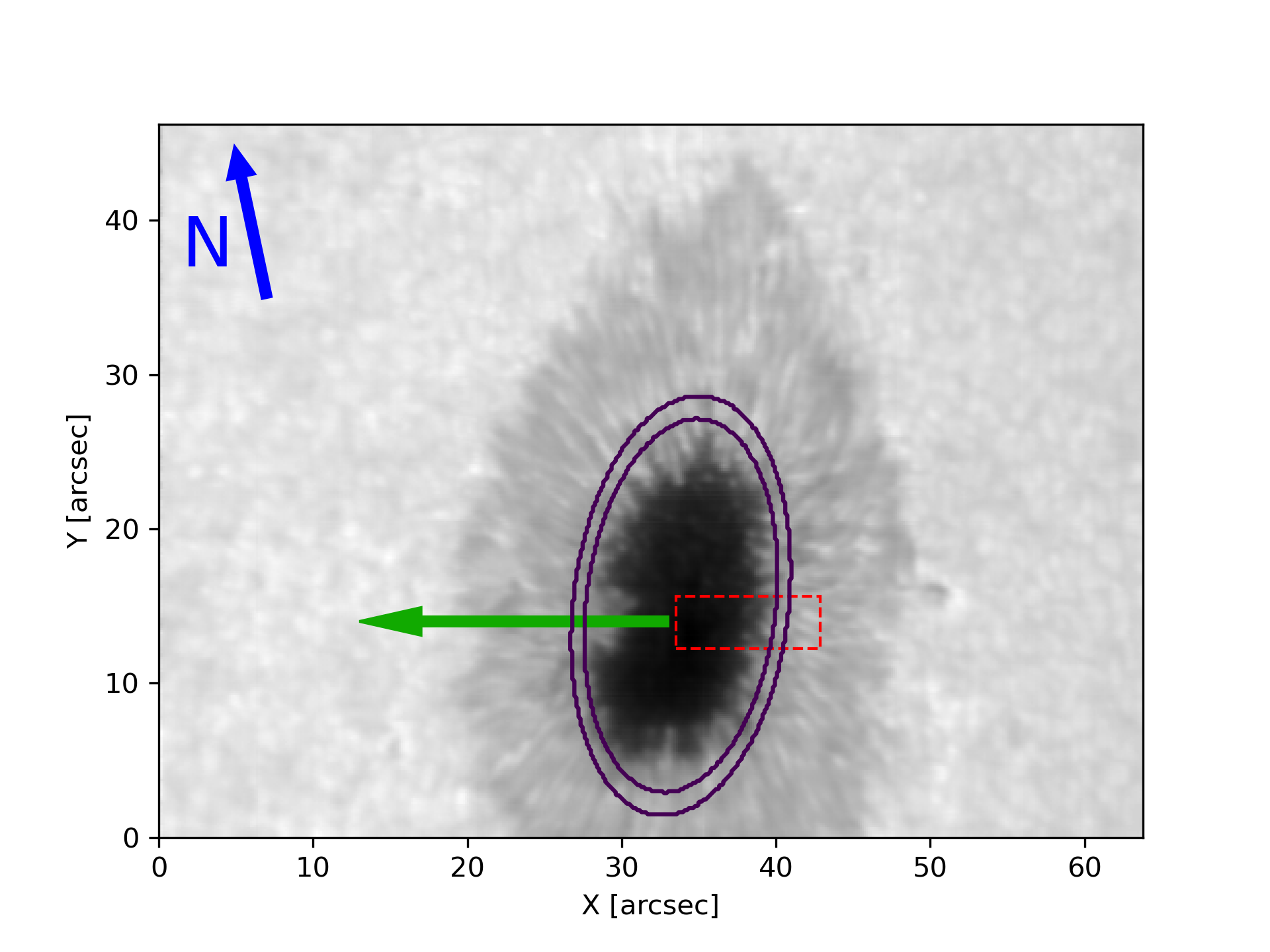}
\caption{Overview of the sunspot NOAA\,12553 in the intensity of the spectral continuum, re-assembled from the scan. 
The x and y axes are given in arcseconds. The green arrow indicates the direction to the solar disk center. 
The blue arrow is pointing north.
The highlighted red rectangle denotes the specific area chosen to calculate the height of spectral lines, as is described in Section \ref{heightOfSpectralLine}. The purple ring limits the area near the Stokes-$V$-zero line chosen to calculate average magnetic field parameters, as is described in Section \ref{Magnetic field}.}
\label{fig:FigGam}
\end{figure}

The magnetic field in sunspots is weaker in the chromosphere than in the photosphere.
Despite the thorough investigation of sunspots, several unsolved questions remain about these phenomena. One of them is the height dependence of their magnetic field (see the review of \citet{Balthasar2018}).
Sunspots are in balance with the atmosphere around them due to the pressure of the stratified gas. The magnetic field in sunspots has its own pressure and reduces the inner gas pressure. To maintain the hydrostatic equilibrium, the magnetic field strength has to decrease with height. 
For this gradient, a controversy arises between studies using various spectral lines and those employing the divB = 0 condition. Multiple spectral lines are formed at different heights and deliver the height dependence of the magnetic field. 
\newline
On the other hand, the divergence of $B$ has to be zero, and the vertical gradient has to balance the horizontal ones, which can be obtained from two-dimensional maps. Deriving the difference of the total magnetic field strengths from two or more spectral lines, values vary between 1.0 to 4.0 $\rm{G\, km}^{-1}$. Following $\rm{div} B = 0$, the values vary between 0.2 and 1.0 $\rm{G\, km}^{-1}$ for different sunspots and spectral lines.
\newline 
To determine these gradients, it is important to get the formation heights of the spectral lines. For the photosphere, typically absorption contribution functions (ACFs) \citep{Grossmann-Doerth1994, Balthasar2008} or response functions \citep{RuizCobo1992} are calculated, based either on a model atmosphere or the inversion of observations. For chromospheric lines such as \ion{He}{i}\,10830\,\AA{}, formation heights are not well known, especially when a magnetic field affects them. For a horizontally extended atmosphere, \citet{Avrett1994} determined heights up to 2000\,km for the helium line. However, the formation height can change a lot if magnetic fields come into the game \citep{Libbrecht2021, Felipe2023}. So far, it remains an open question as to which heights above the sun the helium line is formed at.\newline
In the chromosphere above ARs, fast downflows occur, which exceed the local sound speed \citep{Penn1995, Schmidt2000, Lagg2004}. 
Supersonic downflows in ARs remain a subject of incomplete comprehension, although they were observed many times in solar arch filament systems \citep[AFS,][]{Bruzek1967, Bruzek1969, Lagg2007, GonzalezManrique2018, Manrique2020}. A sunspot is likely accompanied by an arch filament system \citep{Sowmya2022} and these supersonic velocities are consistently observed close to the footpoints \citep[e.g.,][]{Schmidt2000, Lagg2007, Schad2011, GonzalezManrique2018, Yadav2019, Manrique2020}. Various interpretations exist for these supersonic downflows. One of the explanations is that they are a consequence of magnetic flux emergence \citep{GonzalezManrique2018}; another suggestion involves siphon flow attributed to differences in the magnetic field \citep[e.g.,][]{Ruedi1992} between the footprints.\newline
This study investigates the line formation of a \ion{He}{i} 10830\,\AA{} by using dipole approximation and the height dependence of the magnetic field in a sunspot situated in NOAA 12553. Additionally, our focus in this paper is to notice observed supersonic downflows at the sunspot's edge in \ion{He}{i} 10830\,\AA{}. The paper is organized as follows: Observations and data analysis are outlined in Section \ref{section:Observations} and Section \ref{section:Data analysis} respectively. The results are presented in Section \ref{section:Results} and discussed in Section \ref{section:Discussion}.
\begin{figure*}
    \centering
    \begin{subfigure}[b]{\linewidth}
        \centering
        \includegraphics[width=\linewidth]{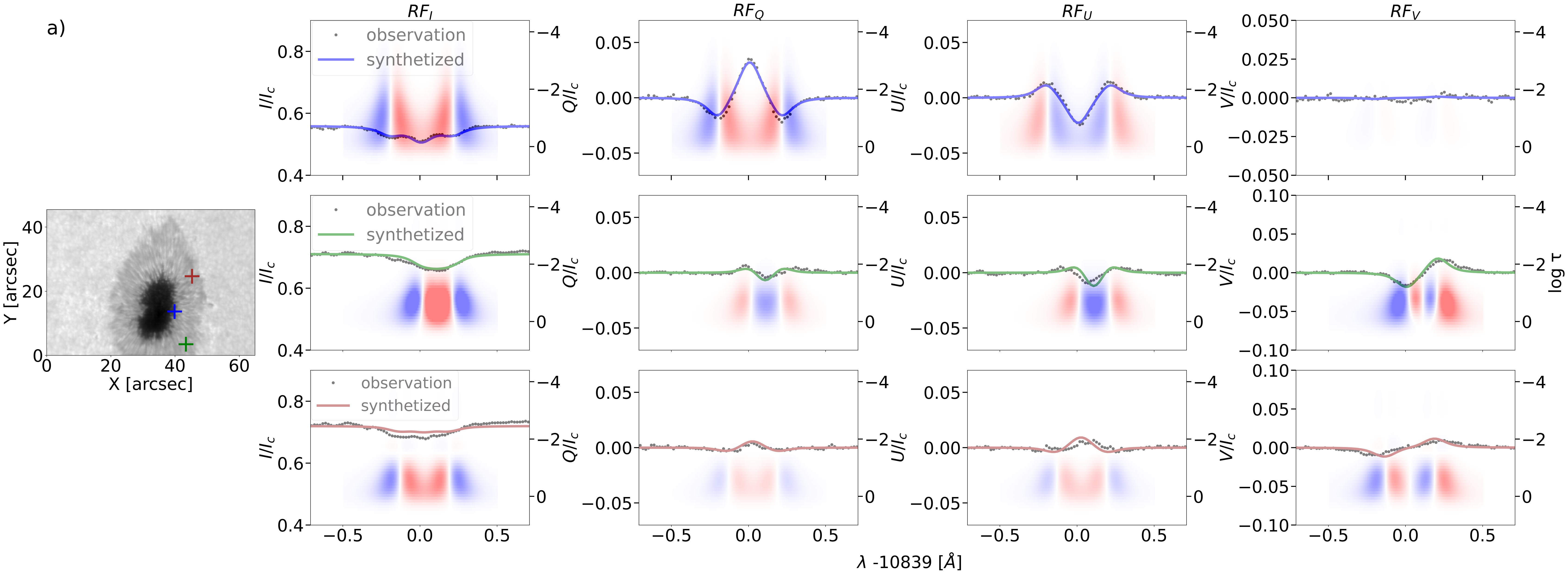}
    \end{subfigure}
    \hfill
    \begin{subfigure}[b]{\linewidth}
        \centering
        \includegraphics[width=\linewidth]{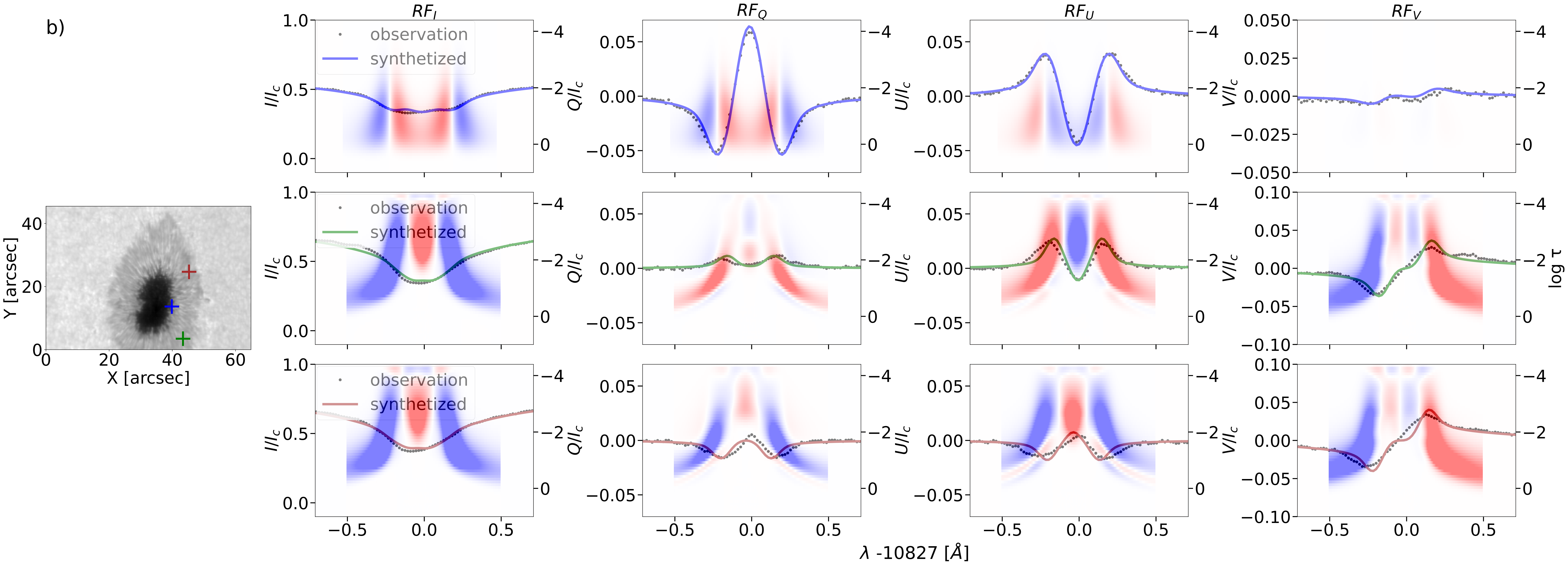}
    \end{subfigure}
    \caption{
    Example of synthesized and observed Stokes profiles $I$, $Q$, $U$ and $V$, along with the corresponding response functions to the magnetic field for a) calcium and b) silicon lines. The calcium intensities ($I/I_\mathrm{c}$) are clipped between 0.4 to 0.9. The log $\tau$ scale is valid for the response functions. The colored crosses mark the locations in the continuum map.}
    \label{fig:sprof}
\end{figure*}
\section{Observations}
\label{section:Observations}
The sunspot, which was part of NOAA 12553, was observed on June 20, 2016 at $x = $ 831\farcs 1 and $y =$ -232\farcs 3,  corresponding to a heliocentric angle of $54^{\circ}$ ($\mu = 0.58$). The observation was acquired with the GREGOR Infrared Spectrograph \citep[GRIS;][]{Collados2012} which is mounted at the 1.5-meter GREGOR solar telescope \citep{Denker2012, schmidt2012}. The AR was observed in spectropolarimetric mode, and the GREGOR polarimetric calibration unit \citep{Hofmann2012} was used to carry out the polarimetric calibration of the data.\newline
We recorded the spectral lines \ion{Si}{i} 10827\,\AA{} (upper photosphere), \ion{He}{i} 10830\,\AA{} (upper chromosphere), and \ion{Ca}{i} 10839\,\AA{} (deeper photosphere). The telescope resolution element is 0\farcs 18 at the used wavelength range under ideal conditions. The slit size is \SI{70}{\micro\meter}, corresponding to 0\farcs 25. We recorded the Stokes parameters of the spectral lines above, between 10:57~UT and 11:23~UT. The observed sunspot was sequentially scanned. The slit orientation (horizontal in Fig~\ref{fig:FigGam}) was perpendicular to the nearby solar limb. The scan consists of 340 slit steps perpendicular to the slit. The step size of the scan was 0\farcs 134, with ten accumulations with single exposure times of \SI{100}{\milli\second}. The slit length was 477 pixels of a size of 0\farcs 136. The field of view (FOV) of the acquired data (shown in Fig.~\ref{fig:FigGam}) has a size of 64\farcs 82 along the slit and 45\farcs 56 in the stepping direction. The spectral sampling is 18 \rm{m\AA{}} per pixel. According to the open data policy, the observation can be freely downloaded from the GRIS data archive.\footnote{https://sdc.leibniz-kis.de} \newline 
The silicon line and the calcium line are both sensitive to the magnetic field (Zeeman triplets), with an effective Landé factor of $g_{eff} = 1.5$. The \ion{He}{i}\,10830\,\AA{} line consists of three spectral lines. The first component appears at 10\,829.09\,\AA{} with $g_{eff} = 2.0$, and is called the blue component. The second component of \ion{He}{i} appears at 10\,830.25\,\AA{} with $g_{eff} = 1.75$, and the third component at 10\,830.34\,\AA{} with  $g_{eff} = 1.25$ is overlapping with the second one, causing them to produce the partial Paschen-Back effect. The latter two components together are called the red component of the helium line.
\begin{figure}[h]
    \centering
\includegraphics[ width= \columnwidth, trim=0cm 0cm 0cm 0.5cm, clip=true]{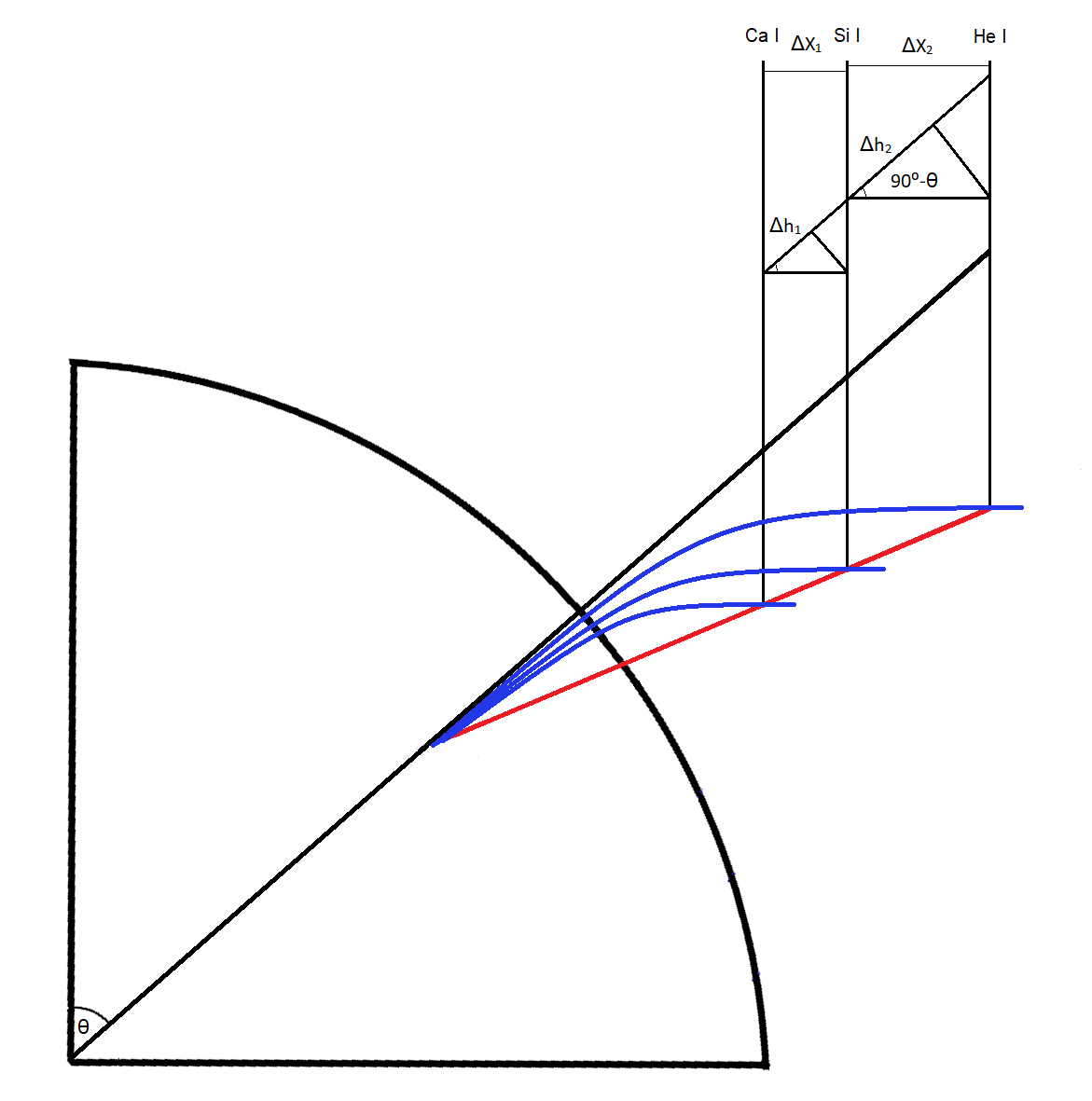}
\caption[width = \columnwidth ]{Schematic sketch of the dipole field in the sunspot magnetic field. The geometry allows one to estimate the formation height of the measured Stokes-$V$ parameter of the spectral lines in the sunspot regions. The line-of-sight vector is characterized by angle $\Theta$. $ \Delta x_{1}$ and $\Delta x_{2}$ are the distances of the Stokes-$V$ zero line between neighboring spectral lines. $ \Delta h_{1}$ and $\Delta h_{2}$ are the real height of the spectral lines.}
\label{fig:FigHeight}
\end{figure}
\begin{figure}[ht]
  \centering
  \includegraphics[width=\linewidth, trim=2.0cm 1.7cm 1cm 13.5cm, clip=true]{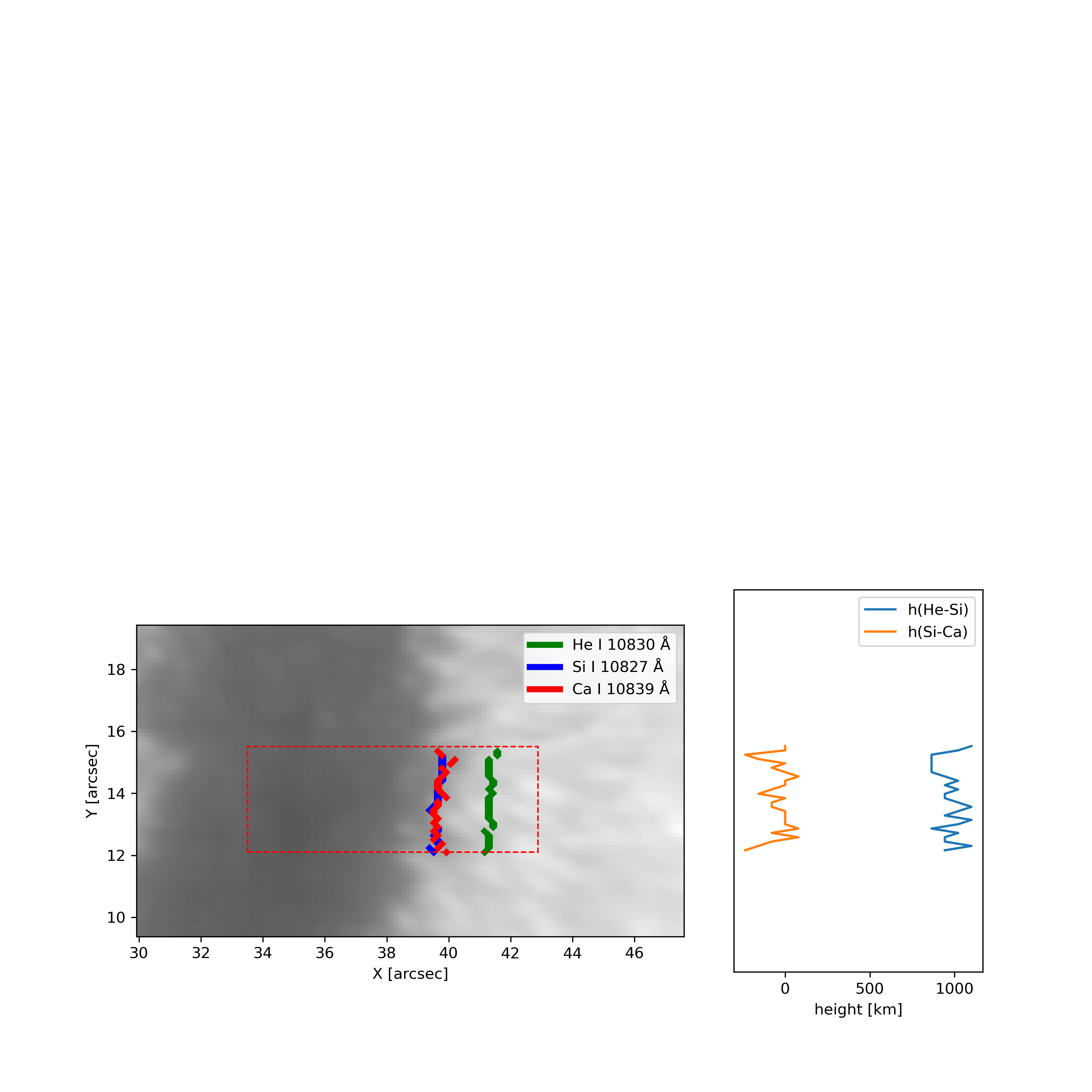}
  \caption{Morphology of the magnetic Stokes-$V$ zero line displayed on the intensity continuum. The individual colors represent values of Stokes-$V$ for \ion{Ca}{i} 10839\, \AA{}, \ion{Si}{i} 10827\,\AA{} and \ion{He}{i} 10830\,\AA{}. The graph shows the calculated height difference for silicon and calcium (orange), and for helium and silicon (blue). }
  \label{fig:figure_heights}
\end{figure}
\section{Data analysis}
\label{section:Data analysis}
In this section, we summarize the data processing performed on observations from the GREGOR telescope. The process includes photometric correction, dark current subtraction, flat-fielding, polarimetric calibration, and cross-talk removal, and was performed with the standard GRIS data reduction software, as has been described by \citet{Collados1999, Collados2003}. The spectrograph still exhibits a continuum intensity variation that has to be corrected. For this purpose as well as for the wavelength calibration, we used the IAG spectral atlas obtained with the Fourier Transform Spectrograph \citep{Reiners2016}. The continuum correction curve was obtained by fitting a polynomial to the ratio of the disk center slope from our observations and the slope of the atlas spectrum. The telluric lines in the observed spectral region were used for the wavelength calibration. The calculated spectral dispersion was 18\,m\AA{} per pixel. Finally, the observed spectrum was normalized to the continuum level of the quiet Sun at the disk center, which was necessary for the inversions.\newline 
\subsection{Inversions} 
\label{inversions}
The Stokes parameters were inverted using the Stokes inversions based on response functions \citep[SIR;][]{RuizCobo1992}. The inversion code provides physical parameters such as the components of the magnetic field vector, the temperature, and the Doppler velocity. We limited the inversions to one node for the Doppler velocity, magnetic field strength, inclination and azimuth, keeping these values height-independent. Only the temperature is height-dependent, with one, three, and four nodes in the different cycles of SIR. 
We used a value of 750\,m\,s$^{-1}$ for the macroturbulence and do not change it during the whole inversion process.
Independent inversions were performed for both the silicon and calcium lines, see plotted Stokes profiles in Fig~\ref{fig:sprof}.
Using SIR, the nodes must be selected following certain rules, and often they do not fit the formation height of the individual lines. Therefore, we prefer separate inversions with one node for the magnetic field vector, which then addresses the formation height of the used spectral lines (see the calculated response functions in Fig~\ref{fig:sprof}). The response functions quantify the sensitivity of an observable (in this case the Stokes parameters) to a specific atmospheric property (in this case the total magnetic field strength, $B_\mathrm{tot}$) in the specific optical depth.\newline  
The GRIS spectropolarimeter enables the study of the chromosphere through the \ion{He}{i} 10830\,\AA{} line, which is formed in the chromosphere. The Stokes parameters of the helium triplet were inverted using the HAnle and ZEeman Light (HAZEL) code. We followed the description of the original paper by \citet{Asensio2008} and the HAZEL manual.\footnote{https://aasensio.github.io/hazel} The inversion conditions were defined in a configuration file and include the determination of Doppler velocity, magnetic field vector, thermal width and optical depth. The inversion process was conducted in two cycles with a single atmospheric component.\newline
To solve the 180$^{\circ}$-ambiguity, we assumed a single azimuth center for the sunspot and a radial structure inside the spot as an ideal configuration. The ideal azimuth vector has a radial structure and the real azimuth values must differ by less than ± $90^{\circ}$ from the ideal case. If abrupt discontinuities in the Cartesian components of the magnetic field remained after this procedure,
we did a manual correction of the azimuth by 180$^{\circ}$ on one side of the discontinuity until we got a continuous magnetic field. A detailed description of this procedure is given by \citet{Balthasar2008}. For helium at the northern tip of the spot, after the described corrections we still encountered discontinuities in the Cartesian components of the magnetic field. We could remove them by changing the azimuth by 90$^\circ$ in some locations. The magnetic field strength is low at these locations, and we cannot exclude that we are dealing here with the ambiguities of the Hanle effect. Therefore, we eventually applied this correction to remove the discontinuities.
The final step was to transform the coordinate system of the magnetic field to the local reference frame and to correct the geometrical foreshortening of the data maps by applying the method described by \citet{Verma2012}.

\subsection{Heights of the \ion{He}{i} 10830\,\AA{}} 
\label{heightOfSpectralLine}
In this subsection, we describe a method of obtaining the formation heights of Stokes-$V$ parameters in the sunspot. Assuming a dipole configuration of the magnetic field as a reasonable first approximation \citep[see][]{LitesSkumanich1990}, the zero line of the Stokes-$V$ parameter allows us to estimate the relative height of the spectral lines. For this, we made use of the dipole property that all field lines cross a selected radial vector from the center of the dipole under the same angle $\alpha$, for which the following equation is valid
\begin{equation}\tan \alpha = 0.5\tan \beta,
\end{equation}
where $\beta$ is the angle between the radial vector and the axis of the dipole \citep[see][]{Kertz1969}. The measured Stokes-$V$ zero-line positions of the three spectral lines were shifted due to their formation height. Along this radial vector, we can determine the distance ratios, and because of the intersection theorem, we have the same ratios along the normal to the solar surface. Here we know the heliocentric angle that corresponds to the zero line in the Stokes-$V$ maps for the different lines. The measured Stokes-$V$ zero-lines were shifted with respect to each other in each spectral line, corresponding to different heights in the solar atmosphere. The valid values are only along the line from the disk center via the sunspot center to the limb (azimuth $\psi = 0^{\circ}$), including an offset of $\pm 5^{\circ}$ to obtain a few more data points, allowing averaging, and $\sin \Delta\psi$ is smaller than 0.1 so that the transversal component is negligible. A simple schematic sketch of the geometry sunspot's magnetic field is displayed in Fig. \ref{fig:FigHeight} for the three spectral lines.
The blue lines represent magnetic field lines for three different spectral lines and the red line represents the radial dipole line that crossed the magnetic field lines. The height distance between the spectral lines was calculated like 
 \begin{equation}
 \Delta h_{1} = \Delta x_{1} \cos{(90^{\circ} - \theta)}; \quad \Delta h_{2} = \Delta x_{2} \cos{(90^{\circ} - \theta)}
 \end{equation}
 \ where $\Delta h_{1}$ is the height distance between \ion{Ca}{i}\,10839\,\AA{} and \ion{Si}{i}\,10827\,\AA{} and $\Delta h_{2}$ is height distance between \ion{Si}{i}\,10827\,\AA{} and \ion{He}{i}\,10830\,\AA{}. The $\Delta x$ values are the corresponding distances of the Stokes-$V$ zero lines. 

\begin{figure*}[hpt]
  \centering
  \includegraphics[width=0.8\linewidth]{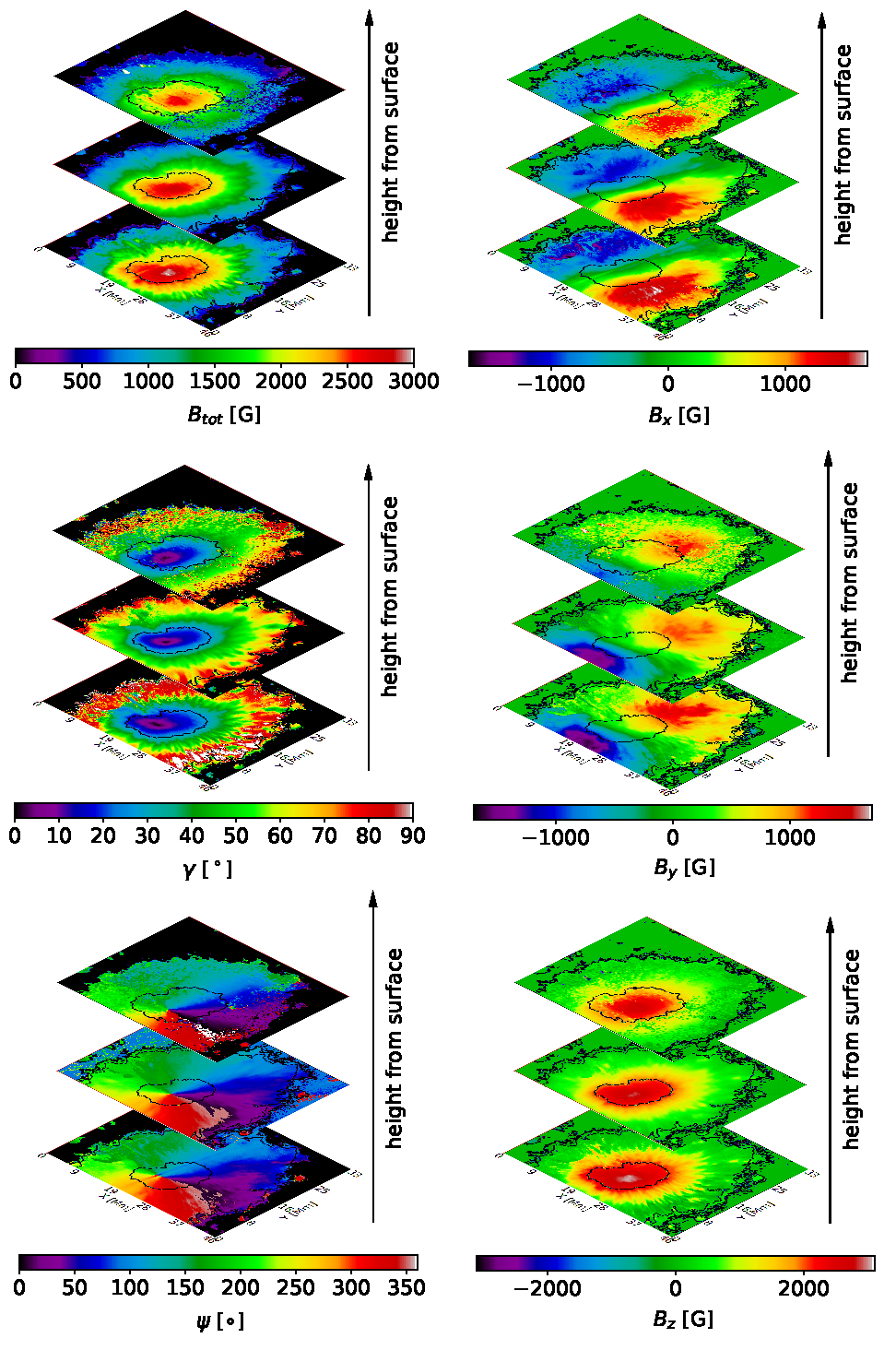}
  \caption{Three layers covered stratification of the magnetic field vector in the solar atmosphere. The displayed heights are not in scale. The left image displays the vector magnetic field in the spherical coordination system (total magnetic field strength, $B_\mathrm{tot}$, inclination, $\gamma$, and azimuth, $\psi$). Inclination values were clipped at 90$^{\circ}$; their maximum is 99$^{\circ}$. The right image displays the vector magnetic field in the Cartesian coordination system ($B_\mathrm{x}$, $B_\mathrm{y}$, $B_\mathrm{z}$). The vector stratification comprises of the \ion{Ca}{i} 10839\,\AA{} and \ion{Si}{i} 10827\,\AA{} for the photosphere and \ion{He}{i} 10830\,\AA{} for the chromosphere. The black contours represent the umbra-penumbral and the penumbra-quiet Sun boundaries.}
  \label{fig:figure_label}
\end{figure*}
   \begin{figure*}[hpt]
   \centering
    \begin{subfigure}[b]{0.4\textwidth}
        \includegraphics[scale=0.5, trim=0.3cm 0.0cm 1.0cm 1.0cm, clip=true]{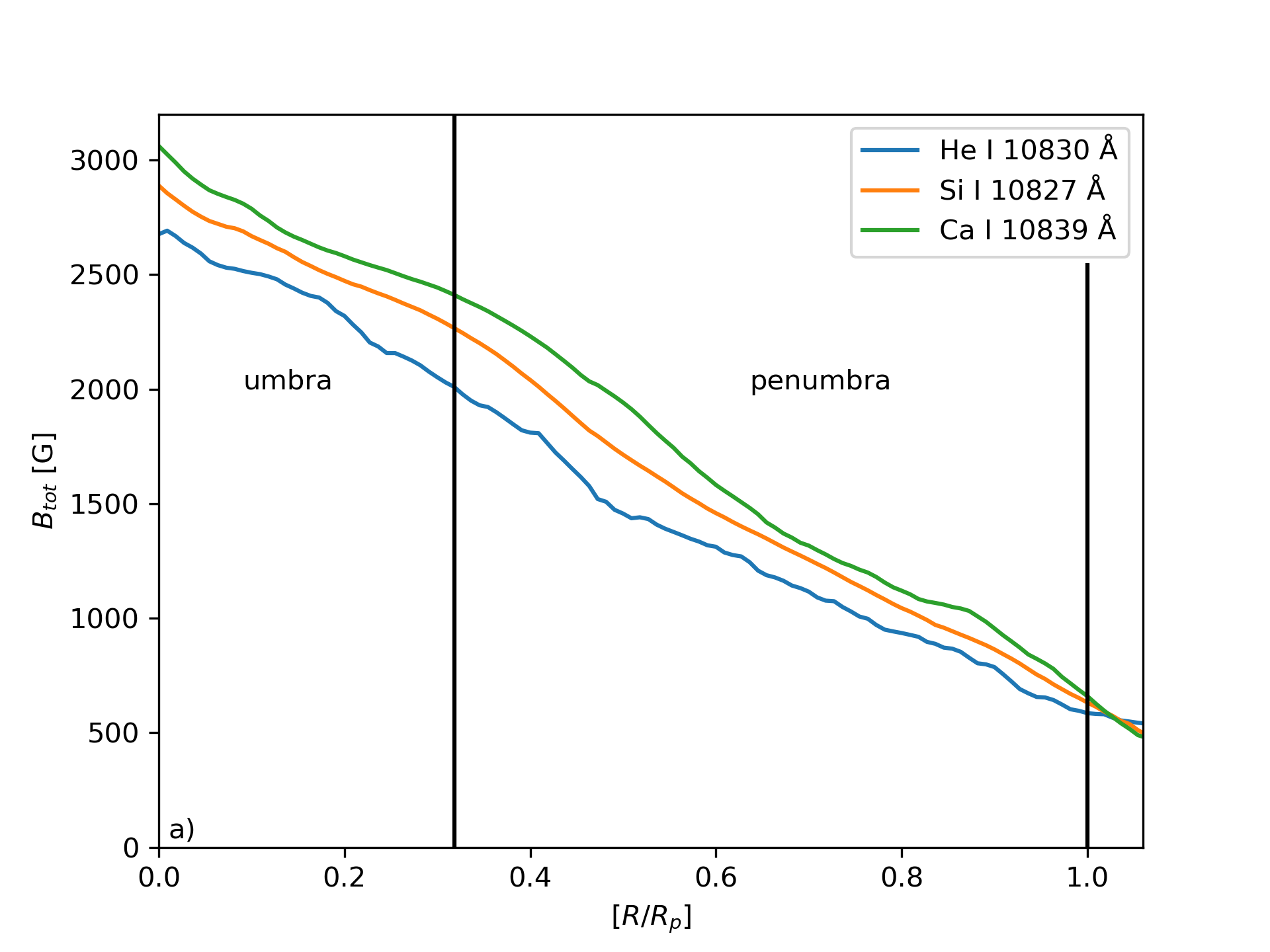}
        \label{fig: a}
    \end{subfigure}
    \begin{subfigure}[b]{0.4\textwidth}
        \includegraphics[scale=0.5, trim=0.3cm 0.0cm 1.0cm 1.0cm, clip=true]{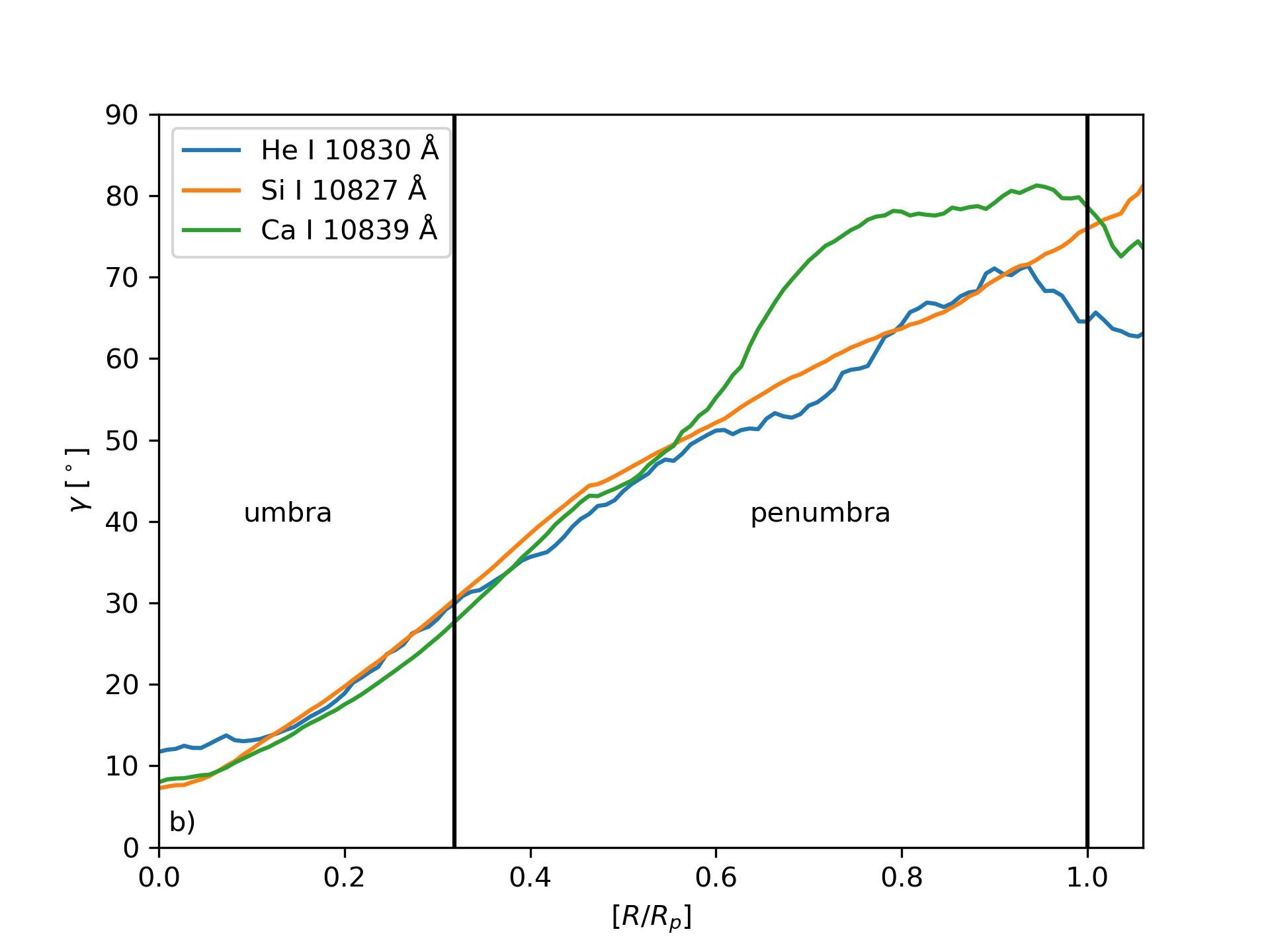}
        \label{fig: b}
    \end{subfigure}
    \begin{subfigure}[b]{0.4\textwidth}
        \includegraphics[scale=0.5, trim=0.3cm 0.3cm 1.0cm 1.0cm, clip=true]{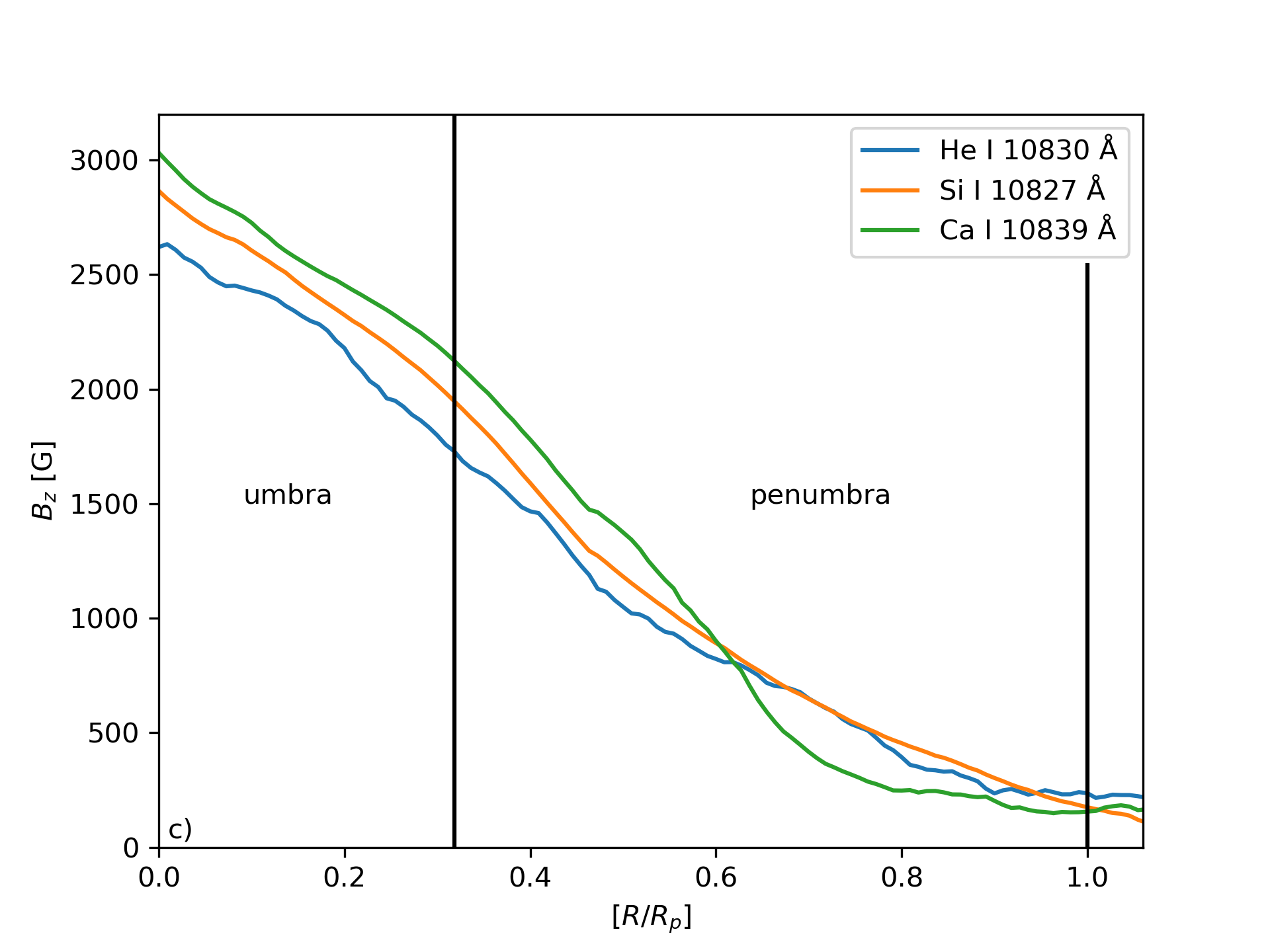}
        \label{fig: c}
    \end{subfigure}
    \begin{subfigure}[b]{0.4\textwidth}
        \includegraphics[scale=0.5, trim=0.3cm 0.3cm 1.0cm 1.0cm, clip=true]{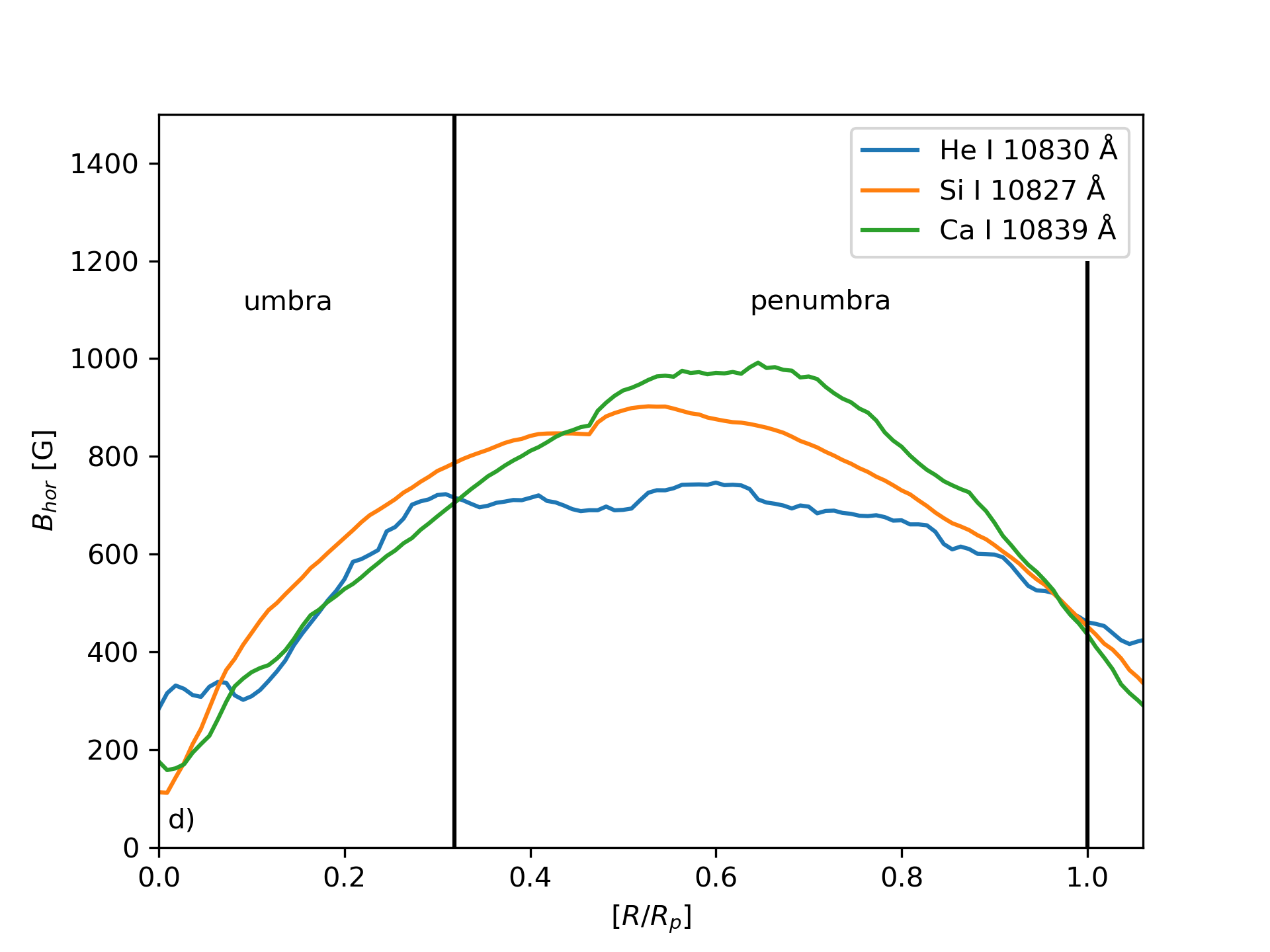}
        \label{fig: d}
    \end{subfigure}
   \caption{Azimuthal averages of the magnetic field components (strength, $B_{tot}$, inclination, $\gamma$, vertical component, $B_\mathrm{z}$, and horizontal component, $B_\mathrm{hor}$) at different distances from the sunspot center are shown. The data are presented for each spectral line in the umbra and penumbra. The x axis represents the distance from the center of the sunspot. The vertical black lines indicate the boundaries between the umbra and the penumbra and between the penumbra and the quiet Sun.}
              \label{fig:graph_rad}
    \end{figure*}
    
\section{Results}
\label{section:Results}

 \begin{figure*}[hpt]
   \centering
    \begin{subfigure}[b]{0.4\textwidth}
        \includegraphics[scale=0.5]{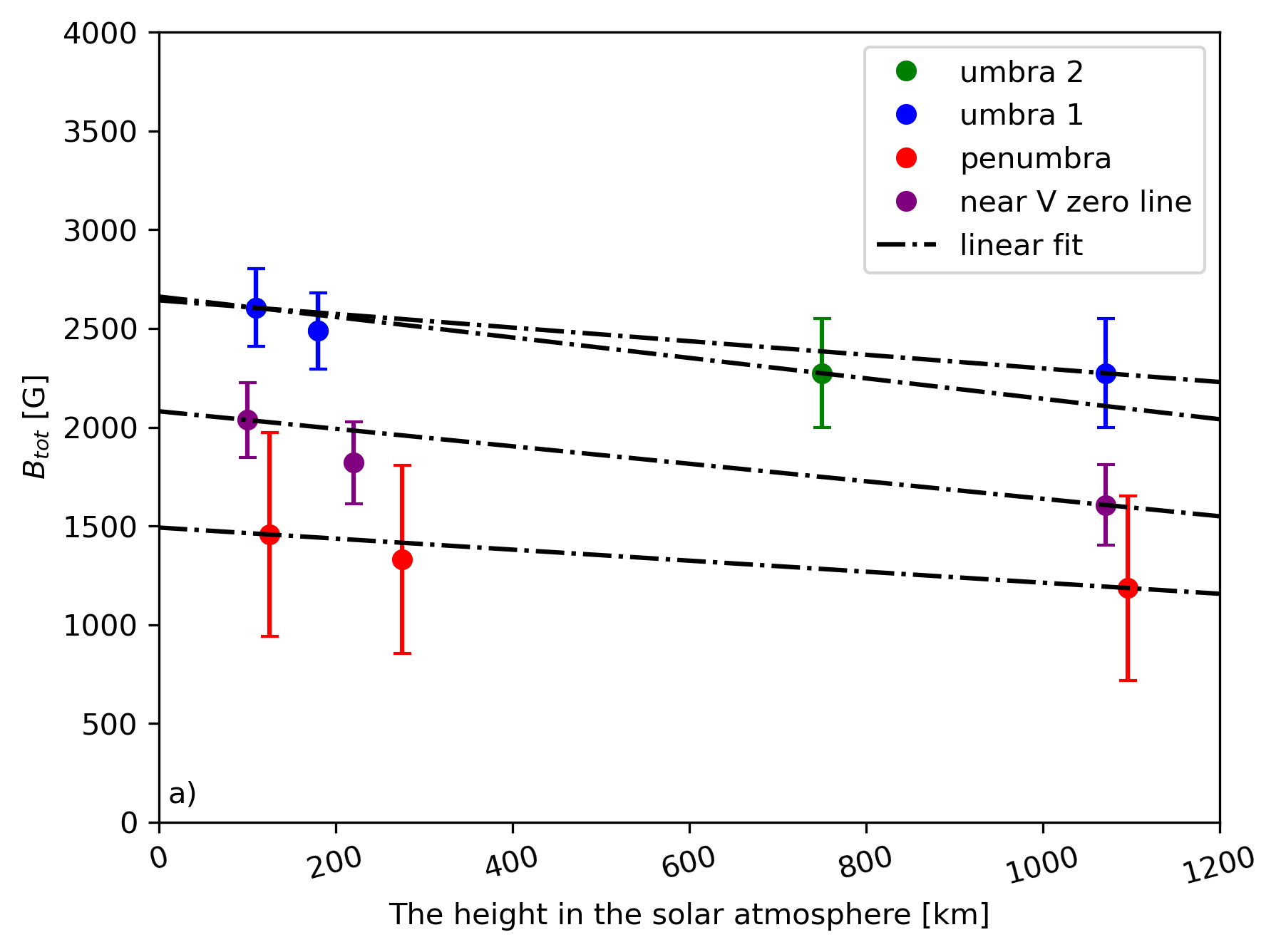}
        \label{figur: a}
    \end{subfigure}
    \begin{subfigure}[b]{0.4\textwidth}
        \includegraphics[scale=0.5]{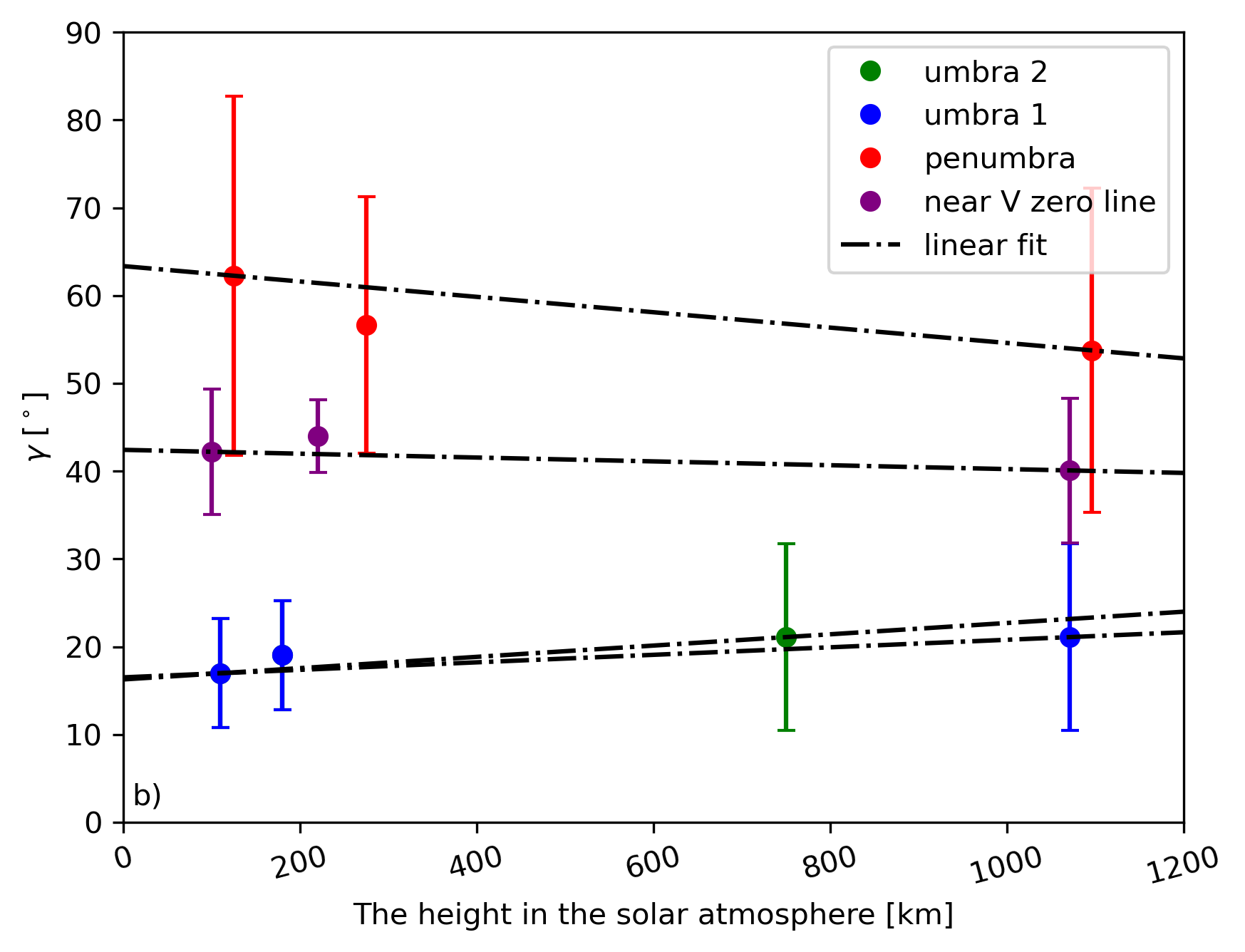}
        \label{figur: b}
    \end{subfigure}
    \begin{subfigure}[b]{0.4\textwidth}
        \includegraphics[scale=0.5]{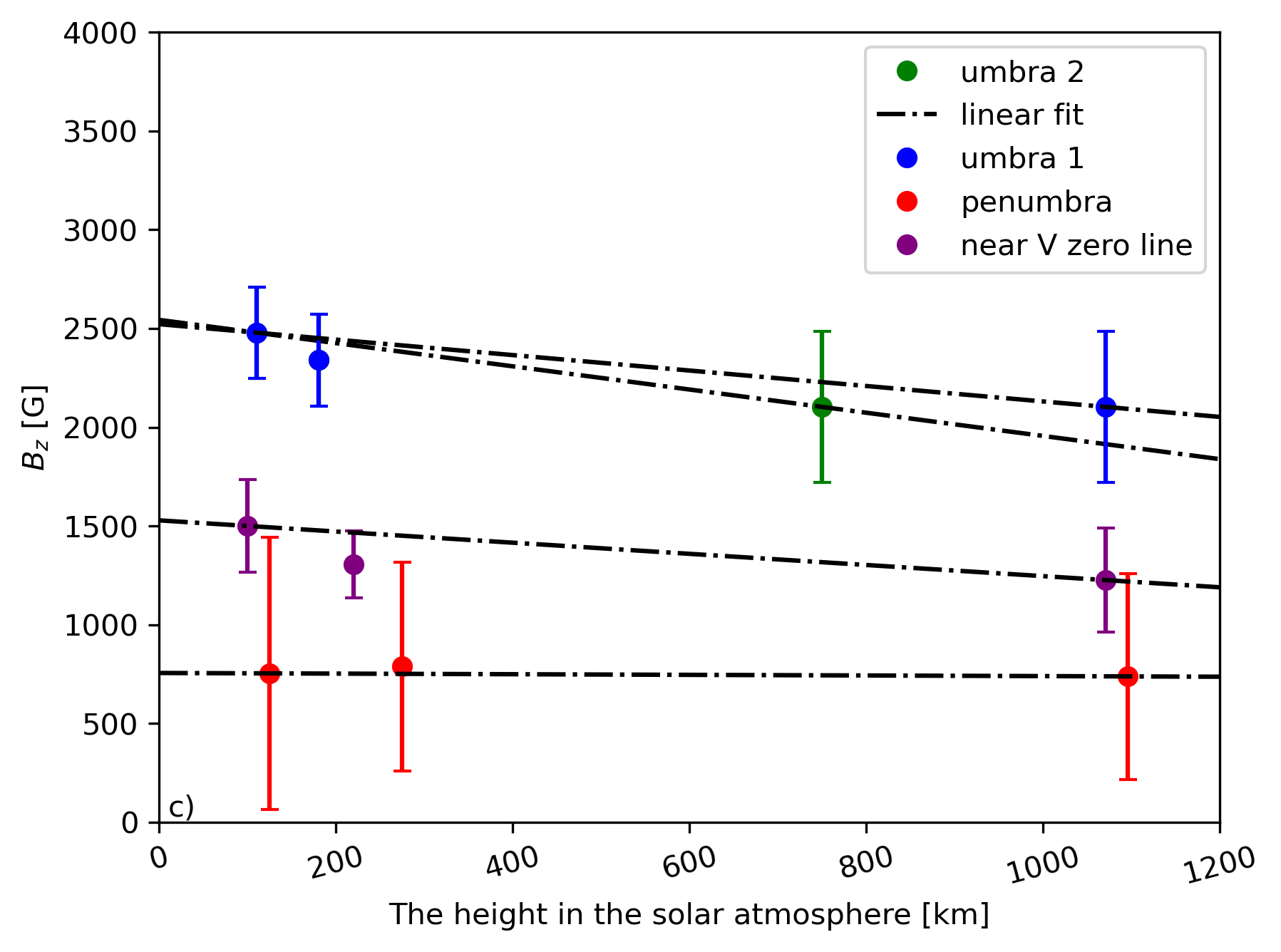}
        \label{figur: c}
    \end{subfigure}
    \begin{subfigure}[b]{0.4\textwidth}
        \includegraphics[scale=0.5]{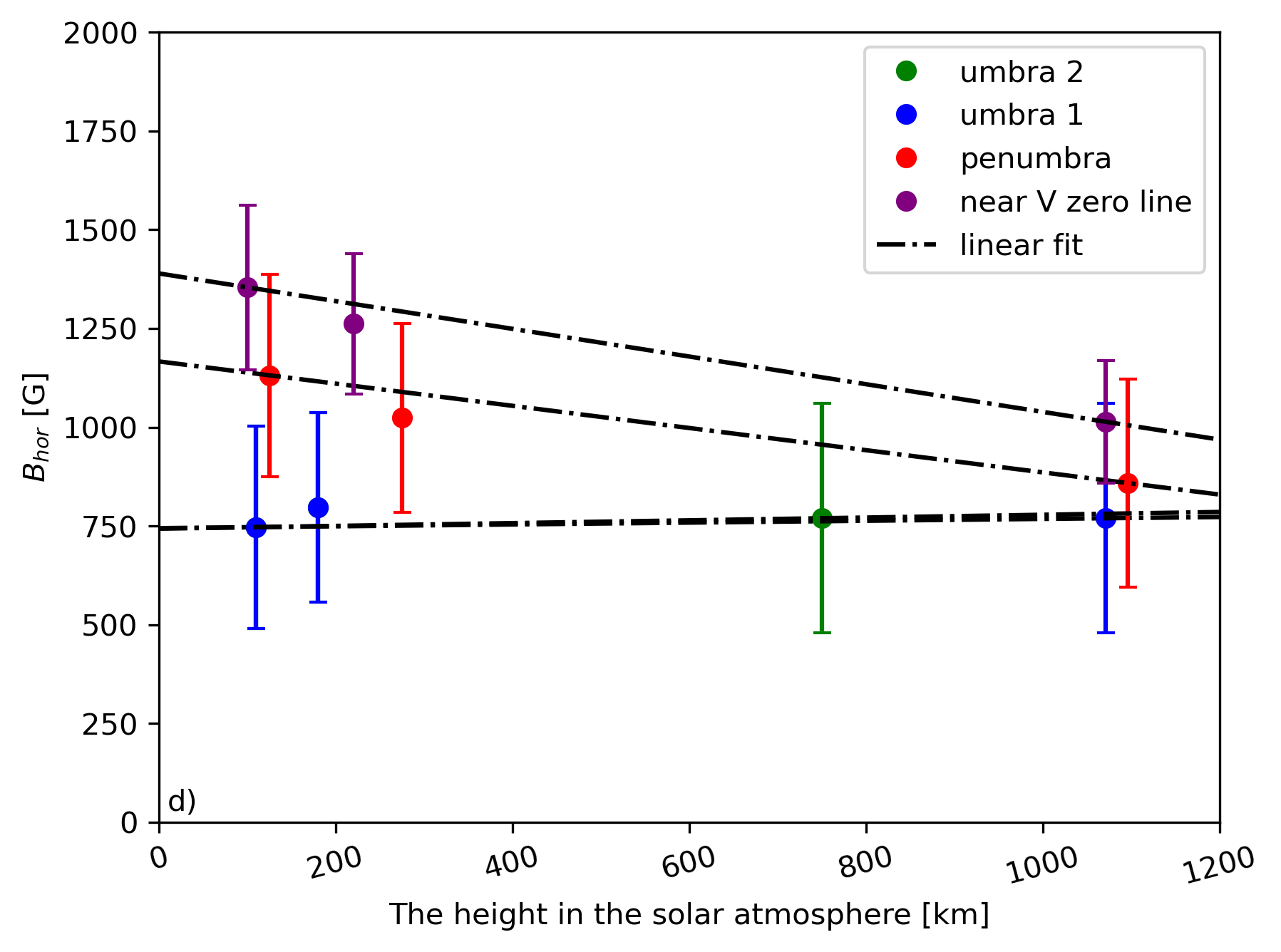}
        \label{figur: d}
    \end{subfigure}
   \caption{Averages of magnetic field parameters (strength, $B_\mathrm{tot}$, inclination, $\gamma$, $B_\mathrm{z}$ and the $B_\mathrm{hor}$ component) and standard deviations of each spectral line in the umbra and penumbra. The x axis displays the height in the solar atmosphere. The height $x=0$ corresponds to $\tau = 1$.  Blue bullets represent the umbra, red ones the penumbra, and violet ones the Stokes-$V$ zero line.}
              \label{fig:Height_av}
    \end{figure*}
\subsection{Magnetic field of the studied sunspot from the photosphere to chromosphere 
\label{Magnetic field}
}

In the first step, we determined the formation height by applying the procedure described in Section \ref{heightOfSpectralLine}. Fig \ref{fig:figure_heights} shows Stokes-$V$ zero lines and the calculated heights of the selected spectral lines. The position of \ion{Si}{i}\,10827\,\AA{} should be located above that of \ion{Ca}{i}\,10839\,\AA{}, but the distance is below the resolution element of the GREGOR telescope at this wavelength; thus, we cannot distinguish between these two lines. We obtain a height difference of 971\,$\pm$ 64 km between the helium and the photospheric lines. The heights of the two photospheric lines were calculated from ACFs for the models M4 by \citet{M4} and T93\_27 by \citet{T93-27} and interpolated between them according to the local conditions \citep[see][]{Balthasar2008}, with 100\,km being obtained for calcium and 220\,km for silicon at the Stokes-$V$ zero line. We prefer ACFs because they provide a more general mean than response functions which depend on the selected physical parameter.\newline
As a result of independently inverted spectral lines (\ion{Ca}{i} 10839\,\AA{}, \ion{Si}{i} 10827\,\AA{}, and \ion{He}{i} 10830\,\AA{}), we obtain the morphology of the magnetic field vector in spherical and Cartesian coordinate systems, covering the magnetic field strength, $B_\mathrm{tot}$, inclination, $\gamma$, azimuth, $\psi$, and the B$_x$, B$_y$, and B$_z$ components of the sunspot. The retrieved results are displayed in Fig.~\ref{fig:figure_label}. Dark contours indicate the umbral boundary at an intensity level of 0.5\,$I_0$, where $I_0$ is the quiet Sun intensity of the local continuum, and the outer penumbral boundary at $B_{\text{hor}} = 490\,\text{G}$, as defined by \citet{Benko2018}. This arrangement of results indicates the three-dimensional structure of the magnetic field in the sunspot from the photosphere to the chromosphere. \newline 
\begin{table*}
  \centering
  
  \begin{tabular}{|r|r|r|r|r|}
    \hline
    \textbf{Units} & \textbf{Umbra 1} & \textbf{Penumbra} & \textbf{near V zero line} & \textbf{Umbra 2}\\
    \hline
    $B_\mathrm{tot}$ [\,$\rm{G\,km}^{-1}$] & $-0.3400$ & $-0.2800$ & $-0.4400$ & $-0.5200$ \\
    \hline
    $\gamma$ ${[^\circ\,\rm{km}^{-1}]} $ & 0.0043 & $-0.0087$ & $-0.0022$ & 0.0064 \\
    \hline
    $B_\mathrm{hor}$ [\,$\rm{G\,km}^{-1}$] & 0.0230 & $-0.2800$ & $-0.3500$ & 0.0350 \\
    \hline
    $B_z$ [\,$\rm{G\,km}^{-1}$] & $-0.3900$ & $-0.0160$ & $-0.2800$ & $-0.5900$ \\
    \hline
  \end{tabular}
\caption{Overview table of the inferred gradients of the magnetic field parameters at the individual parts of the sunspot shown in Fig.\ref{fig:Height_av} The umbra and penumbra values are average values for the whole area. The calculated gradients for umbra 2 were determined with a deeper formation height of the helium line. The values near the Stokes-$V$ zero line are average values from a narrow ring around the whole umbra.
}
  \label{tab:example}
\end{table*}
The spot is a stable one with a positive polarity; thus, the field lines are directed upward. The total magnetic field strength values amount to 3100\,G located in the umbra for \ion{Ca}{i} 10839\,\AA{}. The values decrease from the umbra to the outer penumbra. The inclination angle of the vector magnetic field varies between $0^{\circ}$ to the $100^{\circ}$. The value of $0^{\circ}$ occurs in the center of the umbra, and the inclination increases to $100^{\circ}$ at the outer penumbra. Values above 90$^{\circ}$ indicate a downward-directed magnetic field (see \citet{Westendorp2001b}). The azimuth clearly shows a radial structure across from the umbra to the outer penumbra. The azimuth angle increases continuously in a counterclockwise direction from the zero line. 
\newline
The distribution of the magnetic field vector in the chromosphere is similar to that in the photosphere. The transformed results into the Cartesian coordinate system are displayed in the second column of Fig.~\ref{fig:figure_label}. The vertical component, $B_z$, corresponds to the direction perpendicular to the solar surface, and the horizontal components, B$_x$ and B$_z$, correspond to the plane tangential to the solar surface. B$_x$ is in the direction from the disk center to the solar limb, and B$_y$ is perpendicular to the B$_x$ direction. While B$_z$ dominates in the umbra, B$_x$ and B$_y$ dominate in the penumbra. B$_z$ reaches a maximum value of above 3000\,G at some points in the umbra and decreases toward the outer edge of the penumbra.\newline 
To obtain the radial dependence of the magnetic parameters, we determined mean values along ellipses, omitting those pixels outside the scanned range (see Fig.~\ref{fig:FigGam}) for examples. These ellipses have an increasing distance from the center of the spot. The results are displayed in Fig.~\ref{fig:graph_rad}.
The average values of the vector magnetic field as a function of the radial distance from the sunspot center to the quiet Sun are plotted in Fig.\ref{fig:graph_rad}. The values of $B_\mathrm{tot}$, $\gamma$, $B_\mathrm{z}$, and the $B_\mathrm{hor}$ are displayed in different line colors. The umbra-penumbra boundary is indicated at 0.32 $R/R_p$, where $I$ is 0.5, and the outer penumbra boundary is at 1 $R/R_p$. The inclination of the vector magnetic field is almost vertical in the umbra, but a deviation from the vertical direction appears in the chromosphere because the magnetic field widens. \newline
The inclination increases in the radial direction across the height. The inclination in deep layers (calcium) approaches 90$^{\circ}$ faster than that of other lines between 0.6\,R/R$_p$ and 0.73\,R/R$_p$, and from this point the inclination is almost constant to 1.0\,R/R$_p$. The maximum values are around 80\,$^{\circ}$. For helium, the maximum inclination is 71\,$^{\circ}$ at 0.93\,R/R$_p$ and then it decreases again. Such a decrease might be caused by the fact that the shape of the spot is not perfect, so the ellipses close to the outer penumbral boundary partly cross areas outside it.\newline 
The vertical, $B_\mathrm{z}$, component of the magnetic field vector reaches its maximum values from 2500\,G to 3100\,G in the umbra, and it decreases outward. In the inner penumbra, it reaches values between 1600\,G and 2100\,G and continues to decrease to a value of approximately 200\,G at the outer penumbra. 
The horizontal, $B_\mathrm{hor}$, component of the magnetic field vector has values between 100\,G and 350\,G, increasing toward the penumbra. Values between 650\,G and 1000\,G appear at the inner penumbra. The maximum values are reached at the 0.5\,$R/R_p$ and from this point, the $B_\mathrm{hor}$ values decrease to 400\,G  at the outer penumbra.\newline 
Figure \ref{fig:Height_av} shows the height dependence of a) $B_\mathrm{tot}$, b) $\gamma$, c) $B_\mathrm{z}$, and d) $B_\mathrm{hor}$ for the observed spectral lines. 
The figures present the average values of the selected components for the entire umbra and penumbra. In addition, a narrow ring near the Stokes-$V$-zero line is depicted in Fig.~\ref{fig:FigGam}. Notably, the helium Stokes-$V$-zero line represents a specific region where we encounter challenges in achieving reliable inversions and proper fits for the spectral lines. We suspect that we may be dealing with a situation similar to recent sunspot inversions using HAZEL that were published by \citet{Lindner2023}. It is plausible that the Stokes-$V$ signal is formed at a different atmospheric height than Stokes-$Q$ and Stokes-$U$, as is suggested by \citet{DiazBaso2019a}. 
We applied the height differences determined in Subsection \ref{heightOfSpectralLine} and, in addition, those determined by \citep{Felipe2023}, for an umbral area where strong shocks due to propagating slow magnetoacoustic waves happened. Such shocks can reduce of the response of the helium line to the magnetic field.\newline
In the case of umbra~1, we assumed that the heights obtained in Section \ref{heightOfSpectralLine} are valid all over the spot. The heights derived from ACFs are 110 km for calcium and 180 km for silicon in the umbra. 
For umbra~2, we took a helium height at 750 km in the umbral atmosphere, corresponding to the special case investigated by \citep{Felipe2023} mentioned above.
The calculated heights of the photospheric lines remain the same as in the case of umbra 1. For the penumbra, we obtain from ACFs photospheric heights of 125 km for calcium and 275 km for silicon. As was said above, the formation heights of the photospheric near the Stokes-$V$ zero lines (see Fig.\ref{fig:FigHeight}) are 100\,km and 220\,km, respectively. 
The vertical bars represent the standard deviations from the average value. We applied a linear fit to the data. 
Table\ref{tab:example} shows the gradient values determined by linear fits used in Figure \ref{fig:Height_av}. \newline
The umbral average values of magnetic field strength decrease linearly with increasing height, from 2600\,\rm{G} to 2270\,\rm{G}. The penumbral values of the magnetic field strength also decrease with increasing height. 
The penumbral value decreases from 1460\,\rm{G} to 1180\,\rm{G}. The measured gradient is -0.34\,$\rm{G\,km}^{-1}$ for umbra 1 and -0.52\,$\rm{G\,km}^{-1}$ for umbra 2. A value of -0.28\,$\rm{G\,km}^{-1}$ is obtained for the penumbra. The gradient for umbra 1 is low compared to the values in Table~2 of \citet{Balthasar2018}, while the result for umbra 2 and the penumbral gradient are comparable to the values there.\newline  
Figure \ref{fig:Height_av} b) shows the height dependence of inclination, where the penumbral value decreases in the penumbra, while the umbral value has a tendency to increase with height. The umbral value increases from $17^{\circ}$ to $21^{\circ}$. The penumbral inclination decreases from $62^{\circ}$ to $53^{\circ}$. The measured gradient is 0.0042\,$^{\circ}\,\rm{km}^{-1}$ for the umbra and -0.0087\,$^{\circ}\,\rm{km}^{-1}$ for the penumbra. \newline
Fig. \ref{fig:Height_av} c) shows the $B_{z}$ component of the magnetic field vector. The umbral value shows a clear decrease with increasing height from  2500\,\rm{G} for 110\,\rm{km} to 2100\,\rm{G} in 1070\,\rm{km} over the solar surface. The penumbral value is almost constant. The interval values are between 754\,\rm{G} and 739\,\rm{G}. 
The measured gradient is -0.39\,$\rm{G\,km}^{-1}$ for the umbra and 0.016\,$\rm{G\,km}^{-1}$ for the penumbra.
The last investigated parameter is $B_\mathrm{hor}$, displayed in Fig.~\ref{fig:Height_av} d). The umbral values are almost constant with increasing height. The umbral values vary from 750\,G to 770\,G. The penumbral value clearly decreases depending on the increasing height. The penumbral value is 1130\,G at a height of 110\,\rm{km} and the value decreases to 860\,G for 1071\,km above the solar surface.
The measured gradient is 0.023\,$\rm{G\,km}^{-1}$ for the umbra and -0.28\,$\rm{G\,km}^{-1}$ for the penumbra.
\subsection{Downflows in the \ion{He}{i} 10830\,\AA{}} 
\label{supersonic velocities}
\begin{figure}
   \centering
    \begin{subfigure}[b]{\columnwidth}
    \centering
    \includegraphics[width=\textwidth, trim=0.0cm 2.0cm 2cm 4.0cm,clip=true]{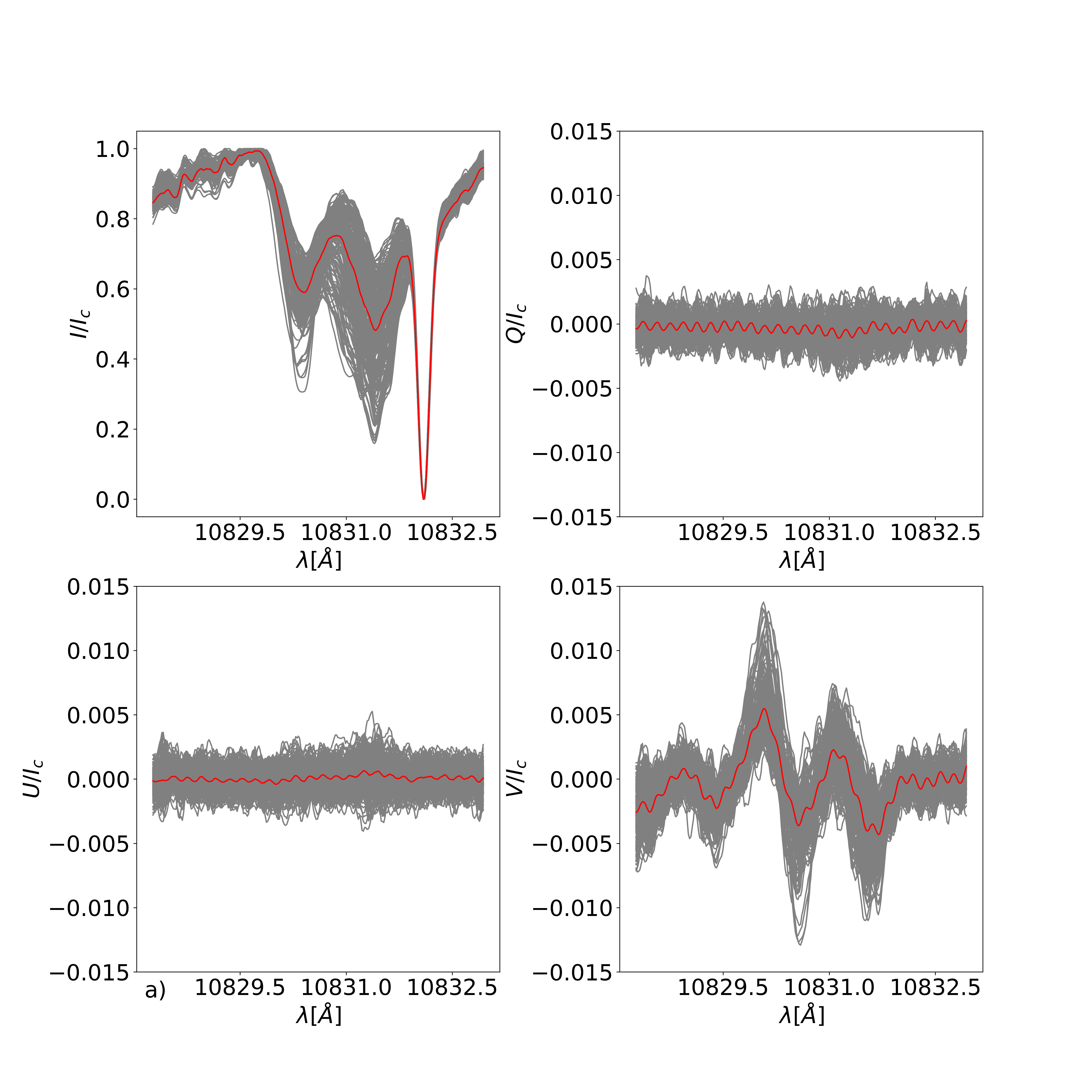}
    \end{subfigure}
    \begin{subfigure}[b]{\columnwidth}
    \centering
    \includegraphics[width=\textwidth, trim=0.0cm 0.0cm 0cm 0.1cm ,clip=true]{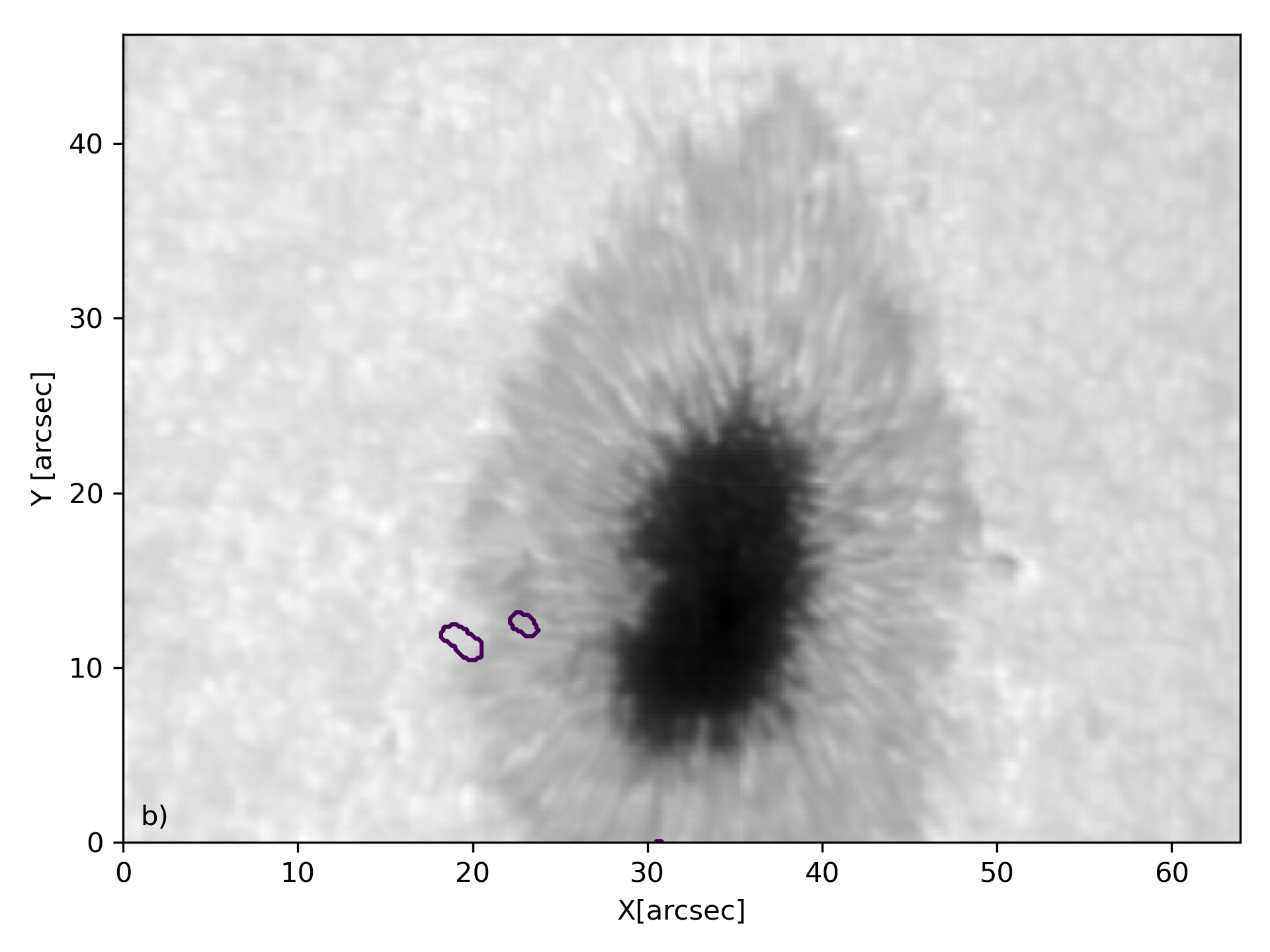}
    \end{subfigure}
    \begin{subfigure}[b]{\columnwidth}
    \centering
    \includegraphics[width=\textwidth, trim=0.0cm 0.0cm 0cm 0.1cm ,clip=true]{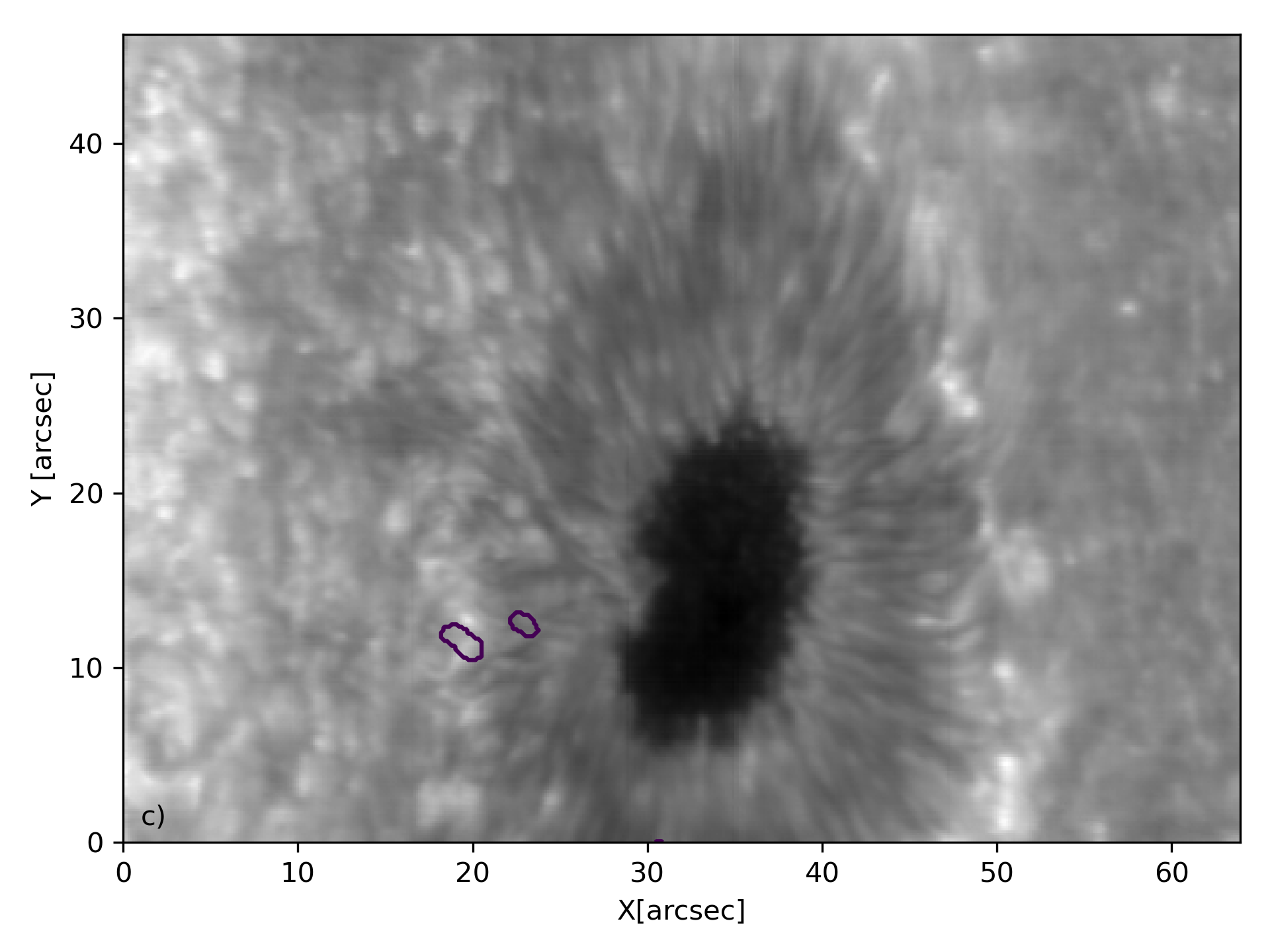}
    \end{subfigure}
    \caption{a) Group of Stokes parameters $I$, $Q$, $U$ and $V$ with a higher Doppler shift located in the outer penumbra.
    b) Location of the group of Stokes parameters with a higher Doppler shift in the continuum.
    c) Location of the group of Stokes parameters with a higher Doppler shift in the silicon core.
    }
    \label{fig:supersonic}
\end{figure}
\begin{figure}[ht]
  \centering
  \includegraphics[width=\linewidth, trim=4.5cm 0.7cm 4cm 0.7cm, clip=true]{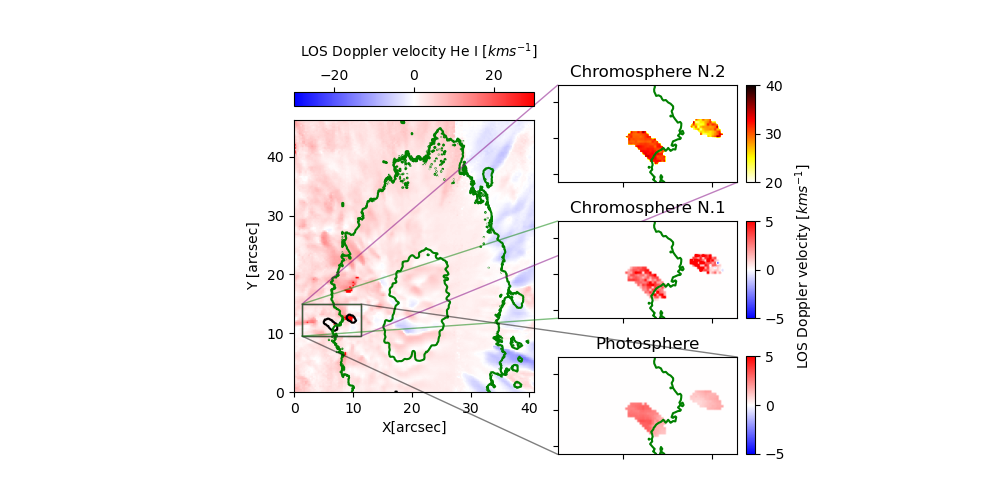}
  \caption{Overview of the helium doppler maps in the sunspot. The left helium Doppler map displays velocities inverted using a one-component atmosphere. The green contours show sunspot intensity boundaries. The purple contour shows areas with high Doppler shifts. The right images show results from inversions with two chromospheric models and one photospheric model. The green line shows the boundary between the penumbra, and the quiet Sun.}
  \label{fig:supersonic_vel}
\end{figure}
The considered area consists of smaller areas where the Stokes-profile shapes are similar due to the fact that the Stokes profiles were formed under similar thermodynamic parameters. At this stage of data processing, we used a special algorithm to classify the observed data before the inversion. While various methods exist for data categorization, choosing this approach proves advantageous in terms of time efficiency. The categorization of the data is helpful for an inversion technique with the goal of determining the physical parameters of the observed atmosphere. Additionally, it serves as a valuable method of quickly scanning and obtaining a comprehensive overview of the entire observed map. One such technique, which we make use of in this paper, is called k-means clustering \citep{macqueen1967}. This technique is often used to group common spectral profiles in large datasets \citep[see, e.g.,][]{Pietarila2007, Viticchie2011, Panos2018, Robustini2019, sainz_dalda2019, Kuckein2020, GonzalezManrique2024}. We also used the MiniBatchKMeans algorithm to reduce the computation time. Both of these algorithms are part of the scikit-learn library of Python. The first step of the algorithm is to assign the initial location of the centroids. The number of clusters is defined by the user.
In our case, we opted for 63 groups to properly cover the diversity of spectra with, all with different characteristic Doppler shifts. It is the minimum number that includes all of the supersonic velocities observed in helium. The creators of the method suggest setting an odd number as it helps one to obtain a clear majority.\newline
The selected profiles are shown in Fig.~\ref{fig:supersonic}\,a) and correspond to clustered Stokes profiles $I$, $Q$, $U$ and $V$ with a higher Doppler shift. Figure~\ref{fig:supersonic}\,b) shows two small patches indicating the location of these profiles in the continuum intensity. The small patches represent areas where the one-component helium inversions do not deliver good results. The right patch is definitely part of the penumbra and it has effects on the behavior of the penumbra and its shape at this place. The left one is just outside the penumbra according to the photospheric boundary. Figure~\ref{fig:supersonic}\,c) complements the continuum image by showing the observed map in the silicon core, where we see penumbra-like structures overlapping with the left patch. These structures are not visible in the continuum map. \newline   
In some special locations, where the red helium component becomes rather asymmetric or even split, an inversion with one atmospheric component is not sufficient. Here we need two atmospheric models \citep[e.g.,][]{Kuckein2020}. The corresponding results are shown in the small images in Fig.~\ref{fig:supersonic_vel}. In the second chromospheric atmosphere, the velocities are supersonic (exceeding the standard sound speed of 10\,$\rm{km\,s}^{-1}$, which is typical for the formation temperature of the \ion{He}{i} 10830\,\AA{} triplet around 8000–10,000 K). They have a tendency to be smaller than 30\,$\rm{km\,s}^{-1}$ inside the penumbra (right patch) and somewhat larger outside that boundary (left patch). The average value of the second helium component is about 30\,$\rm{km\,s}^{-1}$, referred to as the fast velocity component. The average value is 3\,$\rm{km\,s}^{-1}$ $\pm$ 2\,$\rm{km\,s}^{-1}$ for silicon and 2\,$\rm{km\,s}^{-1}$ $\pm$\,0.7\,$\rm{km\,s}^{-1}$ for the first helium component, which is a slow velocity component.\newline
The separated regions evolve under different atmospheric conditions. While the right region is part of the penumbra, the left region is only part of the superpenumbra, where the photospheric Evershed flow does not occur. This area is located at a sharp end of the penumbra structure due to the presence of fast velocities that occur in higher parts of the solar atmosphere and are related to the fast temporal evolution of the sunspot.

\section{Discussion and conclusions}
\label{section:Discussion}
In this analysis, we investigated a sunspot near the limb, observed with GREGOR/GRIS. Not many sunspots that are close to the limb have been studied in detail before. We analyzed the magnetic topology and (thermo)dynamic structure of the sunspot NOAA 12553 at different heights using the spectral lines \ion{Ca}{i} 10839\, \AA{}, \ion{Si}{i} 10827\,\AA{}, and \ion{He}{i} 10830\,\AA{}, which allowed us to determine the stratification from the deep photosphere to the chromosphere. The chromospheric height of \ion{He}{i} 10830\,\AA{} was determined using the properties of a dipole field that we took as a first approximation of the magnetic field of the sunspot. The distance of the Stokes-$V$ zero lines for the different spectral lines was then transferred into different heights. Since the positions of the photospheric lines fall close together within the resolution element of the GREGOR telescope, we cannot use this method to determine the height difference of the photospheric lines. However, we find that the \ion{He}{i} line is formed about 970\,km higher than the photospheric lines at this location. \citet{Felipe2023} find for certain locations in the umbra a formation height of the helium line only 450\,km higher than the silicon line. 
These values are lower than those calculated by \citet{Avrett1994} for plages and the quiet Sun, in which the helium line originates 2000\,km higher than the photosphere. For a quiet Sun simulation, \citet{delaCruzRodr2019} displays the formation height of helium between 1000 and 3000\,km, and at special locations even higher. Nevertheless, the situation in sunspots would be different. 
In our investigation, the mean value of the magnetic field in the umbra amounts to 2600\,G at the height of the calcium line and decreases to a value of 2270\,G at the formation height of the helium line. These values are typical of a large sunspot. The decrease rate is $-0.34$\,$\rm{G\,km}^{-1}$, if we assume a helium formatio n height of 1071\,km. This value is a bit lower than the values listed by \citet{Balthasar2018}, but if we assume a helium formation height of 750\,km, the value is -0.52\,$\rm{G\,km}^{-1}$, a value within the range of \citet{Balthasar2018}.  
The vertical component of the magnetic field strength, $B_\mathrm{z}$, decreases with height in the umbra. The $B_\mathrm{z}$ component amounts to values of 2500\,G in the lowest layer of the photosphere, 2300\,G in the upper part of the photosphere, and only 2100\,G in the lower layer of the chromosphere. The values of the magnetic inclination indicate an almost vertical magnetic field in the umbra; mean values vary between $17^{\circ}$ and $21^{\circ}$. 
As Figs.~\ref{fig:graph_rad} and \ref{fig:Height_av} show, the inclination slightly increases with height, which means that the magnetic field is more inclined in higher layers. This result is in agreement with previous publications. The penumbra of the observed sunspot is usually dominated by a more horizontal magnetic field. The magnetic field strength of the penumbra in the photosphere shows a value of 1450\,$\rm{G}$ for the calcium line and decreases with height to a value of 1180\,$\rm{G}$ for the helium line. Our calculated average value of the penumbral inclination is $62^{\circ}$ for the calcium line, and in the chromosphere the mean inclination is 53$^{\circ}$.  
The value of the inclination angle doesnt not change with increasing height in the inner penumbra, which is in contrast to the results reported by \citet{Borrero2011}, \citet{Balthasar2008}, and \citet{Tiwari2013}, who found that the magnetic field inclination decreases with height in the penumbra. However, our results are consistent with the findings of \citet{SanchezCuberes2005}, who also found that the inclination in the penumbra is independent of height. In the outer penumbra, we find values around 80$^\circ$ for the calcium line indicating an almost horizontal field in deep layers. \newline
The Cartesian components, $B_\mathrm{x}$ and $B_\mathrm{y}$, dominate in the penumbra, in contrast to the umbra, where the $B_\mathrm{z}$ component dominates. This finding corresponds to the magnetic field distribution found in many previous publications. The vertical component of the magnetic induction vector, $B_\mathrm{z}$, decreases with height in the umbra. The $B_\mathrm{z}$ component exceeds values of 2500\,G in the lowest layer of the photosphere, 2340\,G in the upper part of the photosphere, and only 2100\,G in the lower layer of the chromosphere. The vertical component of the magnetic field vector in the penumbra is almost constant with the height, which is in agreement with the variation of the magnetic inclination. The inferred magnetic field component, $B_\mathrm{z}$, falls into the interval from $-1500$\,G to $-1900$\,G for the umbra \citep{Benko2018} and into the interval from $-600$\,G to $-820$\,G in the penumbra. \citet{Balthasar2008} show that the $B_\mathrm{z}$ component reaches smaller values with increasing height in the umbra and that in the outer penumbra the $B_\mathrm{z}$ value increases with height.\newline
The azimuth angle of the magnetic field vector is radially oriented all over the sunspot and changes continuously from $0^{\circ}$ and $360^{\circ}$, as was to be expected. The azimuth does not change with height.\newline
At the sunspot's edge, we observe regions with supersonic velocities at the chromosphere. The chromosphere above ARs is known for hosting fast downflows surpassing the local sound speed \citep[e.g.,][]{Penn1995, Schmidt2000, Solanki2003b, Lagg2004, GonzalezManrique2016, GonzalezManrique2018, Manrique2020}. A statistical study conducted by \citet{Sowmya2022} revealed supersonic downflows in all ARs, comprising 0.2–6.4$\%$ of the observed FOV, suggesting their relatively common occurrence in the upper chromospheres of ARs. Line-of-sight velocities of these supersonic downflows can reach up to 50\,$\rm{km\,s}^{-1}$, with a majority (approximately 92$\%$) coexisting with subsonic flow components \citep{Sowmya2022}. An alternative might be a footpoint of the inverse (superpenumbral) Evershed flow. However, supersonic velocities have only been observed in the transition region or above, but not in chromospheric layers (see the review of \citet{Solanki2003}). A recent investigation by \citet{Romano2023} shows that the superpenumbral filaments end at the outer boundary of the penumbra, but yields velocities clearly below 10\,$\rm{km\,s}^{-1}$ from chromospheric spectral lines. \newline
Future investigations with a new generation of solar telescopes with a higher resolution, for instance the Daniel K. Inouye Solar Telescope \citep{DKIST, Rast2021} or the European Solar Telescope \citep{EST}, will contribute to solving the remaining open questions.

\begin{acknowledgements}
We express our thanks to Dr. A. Asensio Ramos for helpful discussions and his comments to improve the results. We are indebted to Dr. A. Pietrow for carefully reading the manuscript and giving many valuable comments to improve the paper.
The 1.5-meter GREGOR solar telescope was built by a
German consortium under the leadership of the Leibniz-Institut für Sonnenphysik in Freiburg (KIS) with the Leibniz-Institut für Astrophysik Potsdam
(AIP), the Institut für Astrophysik Göttingen (IAG), the Max-Planck-Institut für
Sonnensystemforschung in Göttingen (MPS), and the Instituto de Astrofísica
de Canarias (IAC), and with contributions by the Astronomical Institute of the
Academy of Sciences of the Czech Republic (ASCR). This work
was supported by the project VEGA 2/0043/24. MB is grateful for the support
of the Stefan Schwarz grant of the Slovak Academy of Sciences. SJGM is grateful for the support of the European Research
Council through the grant ERC-2017-CoG771310-PI2FA, and by the
Spanish Ministry of Science and Innovation through the grant PID2021-127487NB-I00. CK acknowledges funding from the European Union's Horizon 2020 research and innovation 
programme under the Marie Sk\l{}odowska-Curie grant agreement No 895955.

\end{acknowledgements}

\bibliographystyle{aa}
\bibliography{references}

\begin{thebibliography}{72}
\expandafter\ifx\csname natexlab\endcsname\relax\def\natexlab#1{#1}\fi

\bibitem[{{Asensio Ramos} {et~al.}(2008){Asensio Ramos}, {Trujillo Bueno}, \&
  {Landi Degl'Innocenti}}]{Asensio2008}
{Asensio Ramos}, A., {Trujillo Bueno}, J., \& {Landi Degl'Innocenti}, E. 2008,
  \apj, 683, 542

\bibitem[{{Avrett} {et~al.}(1994){Avrett}, {Fontenla}, \&
  {Loeser}}]{Avrett1994}
{Avrett}, E.~H., {Fontenla}, J.~M., \& {Loeser}, R. 1994, in In: Infrared Solar
  Physics, Vol. 154, 35--47

\bibitem[{{Balthasar}(2018)}]{Balthasar2018}
{Balthasar}, H. 2018, \solphys, 293, 120

\bibitem[{{Balthasar} {et~al.}(2013){Balthasar}, {Beck}, {G{\"o}m{\"o}ry},
  {Muglach}, {Puschmann}, {Shimizu}, \& {Verma}}]{Balthasar2013}
{Balthasar}, H., {Beck}, C., {G{\"o}m{\"o}ry}, P., {et~al.} 2013, Central
  European Astrophysical Bulletin, 37, 435

\bibitem[{{Balthasar} \& {G{\"o}m{\"o}ry}(2008)}]{Balthasar2008}
{Balthasar}, H. \& {G{\"o}m{\"o}ry}, P. 2008, \aap, 488, 1085

\bibitem[{{Bellot Rubio} {et~al.}(2004){Bellot Rubio}, {Balthasar}, \&
  {Collados}}]{Bellot2004}
{Bellot Rubio}, L.~R., {Balthasar}, H., \& {Collados}, M. 2004, \aap, 427, 319

\bibitem[{{Benko} {et~al.}(2018){Benko}, {Gonz{\'a}lez Manrique}, {Balthasar},
  {G{\"o}m{\"o}ry}, {Kuckein}, \& {Jur{\v{c}}{\'a}k}}]{Benko2018}
{Benko}, M., {Gonz{\'a}lez Manrique}, S.~J., {Balthasar}, H., {et~al.} 2018,
  \aap, 620, A191

\bibitem[{{Benko} {et~al.}(2021){Benko}, {Gonz{\'a}lez Manrique}, {Balthasar},
  {G{\"o}m{\"o}ry}, {Kuckein}, \& {Jur{\v{c}}{\'a}k}}]{Benko2021}
{Benko}, M., {Gonz{\'a}lez Manrique}, S.~J., {Balthasar}, H., {et~al.} 2021,
  \aap, 652, C7

\bibitem[{{Bommier}(2013)}]{Bommier2013}
{Bommier}, V. 2013, Physics Research International, 2013, 195403

\bibitem[{{Borrero} \& {Ichimoto}(2011)}]{Borrero2011}
{Borrero}, J.~M. \& {Ichimoto}, K. 2011, Living Reviews in Solar Physics, 8, 4

\bibitem[{{Bruzek}(1967)}]{Bruzek1967}
{Bruzek}, A. 1967, \solphys, 2, 451

\bibitem[{{Bruzek}(1969)}]{Bruzek1969}
{Bruzek}, A. 1969, \solphys, 8, 29

\bibitem[{{Carlsson} {et~al.}(2019){Carlsson}, {De Pontieu}, \&
  {Hansteen}}]{Carlsson2019}
{Carlsson}, M., {De Pontieu}, B., \& {Hansteen}, V.~H. 2019, \araa, 57, 189

\bibitem[{{Collados}(1999)}]{Collados1999}
{Collados}, M. 1999, in Astronomical Society of the Pacific Conference Series,
  Vol. 184, Third Advances in Solar Physics Euroconference: Magnetic Fields and
  Oscillations, ed. B.~{Schmieder}, A.~{Hofmann}, \& J.~{Staude}, 3--22

\bibitem[{{Collados} {et~al.}(2012){Collados}, {L{\'o}pez}, {P{\'a}ez},
  {Hern{\'a}ndez}, {Reyes}, {Calcines}, {Ballesteros}, {D{\'\i}az}, {Denker},
  {Lagg}, {Schlichenmaier}, {Schmidt}, {Solanki}, {Strassmeier}, {von der
  L{\"u}he}, \& {Volkmer}}]{Collados2012}
{Collados}, M., {L{\'o}pez}, R., {P{\'a}ez}, E., {et~al.} 2012, Astronomische
  Nachrichten, 333, 872

\bibitem[{{Collados}(2003)}]{Collados2003}
{Collados}, M.~V. 2003, in Society of Photo-Optical Instrumentation Engineers
  (SPIE) Conference Series, Vol. 4843, Polarimetry in Astronomy, ed.
  S.~{Fineschi}, 55--65

\bibitem[{{de la Cruz Rodr{\'\i}guez} {et~al.}(2019){de la Cruz
  Rodr{\'\i}guez}, {Leenaarts}, {Danilovic}, \&
  {Uitenbroek}}]{delaCruzRodr2019}
{de la Cruz Rodr{\'\i}guez}, J., {Leenaarts}, J., {Danilovic}, S., \&
  {Uitenbroek}, H. 2019, \aap, 623, A74

\bibitem[{{Denker} {et~al.}(2012){Denker}, {Lagg}, {Puschmann}, {Schmidt},
  {Schmidt}, {Sobotka}, {Soltau}, {Strassmeier}, {Volkmer}, {von der Luehe},
  {Solanki}, {Balthasar}, {Bello Gonzalez}, {Berkefeld}, {Collados Vera},
  {Hofmann}, \& {Kneer}}]{Denker2012}
{Denker}, C., {Lagg}, A., {Puschmann}, K.~G., {et~al.} 2012, IAU Special
  Session, 6, E2.03

\bibitem[{{D{\'\i}az Baso} {et~al.}(2019){D{\'\i}az Baso}, {Mart{\'\i}nez
  Gonz{\'a}lez}, \& {Asensio Ramos}}]{DiazBaso2019a}
{D{\'\i}az Baso}, C.~J., {Mart{\'\i}nez Gonz{\'a}lez}, M.~J., \& {Asensio
  Ramos}, A. 2019, \aap, 625, A128

\bibitem[{{Felipe} {et~al.}(2023){Felipe}, {Gonz{\'a}lez Manrique},
  {Sangeetha}, \& {Asensio Ramos}}]{Felipe2023}
{Felipe}, T., {Gonz{\'a}lez Manrique}, S.~J., {Sangeetha}, C.~R., \& {Asensio
  Ramos}, A. 2023, \aap, 676, A77

\bibitem[{{Gonz{\'a}lez Manrique} {et~al.}(2024){Gonz{\'a}lez Manrique},
  {Khomenko}, {Collados}, {Kuckein}, {Felipe}, \&
  {G{\"o}m{\"o}ry}}]{GonzalezManrique2024}
{Gonz{\'a}lez Manrique}, S.~J., {Khomenko}, E., {Collados}, M., {et~al.} 2024,
  \aap, 681, A114

\bibitem[{{Gonz{\'a}lez Manrique} {et~al.}(2018){Gonz{\'a}lez Manrique},
  {Kuckein}, {Collados}, {Denker}, {Solanki}, {G{\"o}m{\"o}ry}, {Verma},
  {Balthasar}, {Lagg}, \& {Diercke}}]{GonzalezManrique2018}
{Gonz{\'a}lez Manrique}, S.~J., {Kuckein}, C., {Collados}, M., {et~al.} 2018,
  \aap, 617, A55

\bibitem[{{Gonz{\'a}lez Manrique} {et~al.}(2016){Gonz{\'a}lez Manrique},
  {Kuckein}, {Pastor Yabar}, {Collados}, {Denker}, {Fischer}, {G{\"o}m{\"o}ry},
  {Diercke}, {Bello Gonz{\'a}lez}, {Schlichenmaier}, {Balthasar}, {Berkefeld},
  {Feller}, {Hoch}, {Hofmann}, {Kneer}, {Lagg}, {Nicklas}, {Orozco Su{\'a}rez},
  {Schmidt}, {Schmidt}, {Sigwarth}, {Sobotka}, {Solanki}, {Soltau}, {Staude},
  {Strassmeier}, {Verma}, {Volkmer}, {von der L{\"u}he}, \&
  {Waldmann}}]{GonzalezManrique2016}
{Gonz{\'a}lez Manrique}, S.~J., {Kuckein}, C., {Pastor Yabar}, A., {et~al.}
  2016, Astronomische Nachrichten, 337, 1057

\bibitem[{{Gonz{\'a}lez Manrique} {et~al.}(2020){Gonz{\'a}lez Manrique},
  {Kuckein}, {Pastor Yabar}, {Diercke}, {Collados}, {G{\"o}m{\"o}ry}, {Zhong},
  {Hou}, \& {Denker}}]{Manrique2020}
{Gonz{\'a}lez Manrique}, S.~J., {Kuckein}, C., {Pastor Yabar}, A., {et~al.}
  2020, \apj, 890, 82

\bibitem[{{Grossmann-Doerth}(1994)}]{Grossmann-Doerth1994}
{Grossmann-Doerth}, U. 1994, \aap, 285, 1012

\bibitem[{{Hale}(1908)}]{Hale1908}
{Hale}, G.~E. 1908, \apj, 28, 315

\bibitem[{Hofmann {et~al.}(2012)Hofmann, Arlt, Balthasar, Bauer, Bittner,
  Paschke, Popow, Rendtel, Soltau, \& Waldmann}]{Hofmann2012}
Hofmann, A., Arlt, K., Balthasar, H., {et~al.} 2012, Astronomische Nachrichten,
  333, 854

\bibitem[{{Joshi} {et~al.}(2017){Joshi}, {Lagg}, {Hirzberger}, {Solanki}, \&
  {Tiwari}}]{Joshi2017}
{Joshi}, J., {Lagg}, A., {Hirzberger}, J., {Solanki}, S.~K., \& {Tiwari}, S.~K.
  2017, \aap, 599, A35

\bibitem[{{Joshi} {et~al.}(2016){Joshi}, {Lagg}, {Solanki}, {Feller},
  {Collados}, {Orozco Su{\'a}rez}, {Schlichenmaier}, {Franz}, {Balthasar},
  {Denker}, {Berkefeld}, {Hofmann}, {Kiess}, {Nicklas}, {Pastor Yabar},
  {Rezaei}, {Schmidt}, {Schmidt}, {Sobotka}, {Soltau}, {Staude}, {Strassmeier},
  {Volkmer}, {von der L{\"u}he}, \& {Waldmann}}]{Joshi2016}
{Joshi}, J., {Lagg}, A., {Solanki}, S.~K., {et~al.} 2016, \aap, 596, A8

\bibitem[{{Jur{\v{c}}{\'a}k}(2011)}]{Jurcak2011}
{Jur{\v{c}}{\'a}k}, J. 2011, \aap, 531, A118

\bibitem[{Kertz(1969)}]{Kertz1969}
Kertz, W. 1969, BI-Hochschultaschenb{\"u}cher, Vol. 275, Einf{\"u}hrung in die
  Geophysik (Bibliographisches Institut (Mannheim)), 129

\bibitem[{{Kollatschny} {et~al.}(1980){Kollatschny}, {Wiehr}, {Stellmacher}, \&
  {Falipou}}]{M4}
{Kollatschny}, W., {Wiehr}, E., {Stellmacher}, G., \& {Falipou}, M.~A. 1980,
  \aap, 86, 245

\bibitem[{{Kuckein} {et~al.}(2020){Kuckein}, {Gonz{\'a}lez Manrique}, {Kleint},
  \& {Asensio Ramos}}]{Kuckein2020}
{Kuckein}, C., {Gonz{\'a}lez Manrique}, S.~J., {Kleint}, L., \& {Asensio
  Ramos}, A. 2020, \aap, 640, A71

\bibitem[{{Lagg} {et~al.}(2004){Lagg}, {Woch}, {Krupp}, \&
  {Solanki}}]{Lagg2004}
{Lagg}, A., {Woch}, J., {Krupp}, N., \& {Solanki}, S.~K. 2004, \aap, 414, 1109

\bibitem[{{Lagg} {et~al.}(2007){Lagg}, {Woch}, {Solanki}, \&
  {Krupp}}]{Lagg2007}
{Lagg}, A., {Woch}, J., {Solanki}, S.~K., \& {Krupp}, N. 2007, \aap, 462, 1147

\bibitem[{{Libbrecht} {et~al.}(2021){Libbrecht}, {Bj{\o}rgen}, {Leenaarts}, {de
  la Cruz Rodr{\'\i}guez}, {Hansteen}, \& {Joshi}}]{Libbrecht2021}
{Libbrecht}, T., {Bj{\o}rgen}, J.~P., {Leenaarts}, J., {et~al.} 2021, \aap,
  652, A146

\bibitem[{{Lindner} {et~al.}(2023){Lindner}, {Kuckein}, {Gonz{\'a}lez
  Manrique}, {Bello Gonz{\'a}lez}, {Kleint}, \& {Berkefeld}}]{Lindner2023}
{Lindner}, P., {Kuckein}, C., {Gonz{\'a}lez Manrique}, S.~J., {et~al.} 2023,
  \aap, 673, A64

\bibitem[{{Lites} \& {Skumanich}(1990)}]{LitesSkumanich1990}
{Lites}, B.~W. \& {Skumanich}, A. 1990, \apj, 348, 747

\bibitem[{MacQueen(1967)}]{macqueen1967}
MacQueen, J.~B. 1967, Some Methods for Classification and Analysis of
  Multivariate Observations, Vol.~1 (Berkeley: University of California Press),
  281--297

\bibitem[{{Merenda} {et~al.}(2011){Merenda}, {Lagg}, \&
  {Solanki}}]{Merenda2011}
{Merenda}, L., {Lagg}, A., \& {Solanki}, S.~K. 2011, \aap, 532, A63

\bibitem[{{Moran} {et~al.}(2000){Moran}, {Deming}, {Jennings}, \&
  {McCabe}}]{Moran2000}
{Moran}, T., {Deming}, D., {Jennings}, D.~E., \& {McCabe}, G. 2000, \apj, 533,
  1035

\bibitem[{{Orozco Suarez} {et~al.}(2005){Orozco Suarez}, {Lagg}, \&
  {Solanki}}]{Orozco2005}
{Orozco Suarez}, D., {Lagg}, A., \& {Solanki}, S.~K. 2005, in ESA Special
  Publication, Vol. 596, Chromospheric and Coronal Magnetic Fields, ed. D.~E.
  {Innes}, A.~{Lagg}, \& S.~A. {Solanki}, 59.1

\bibitem[{{Panos} {et~al.}(2018){Panos}, {Kleint}, {Huwyler}, {Krucker},
  {Melchior}, {Ullmann}, \& {Voloshynovskiy}}]{Panos2018}
{Panos}, B., {Kleint}, L., {Huwyler}, C., {et~al.} 2018, \apj, 861, 62

\bibitem[{{Penn} \& {Kuhn}(1995)}]{Penn1995}
{Penn}, M.~J. \& {Kuhn}, J.~R. 1995, \apjl, 441, L51

\bibitem[{{Pietarila} {et~al.}(2007){Pietarila}, {Socas-Navarro}, \&
  {Bogdan}}]{Pietarila2007}
{Pietarila}, A., {Socas-Navarro}, H., \& {Bogdan}, T. 2007, \apj, 663, 1386

\bibitem[{{Quintero Noda} {et~al.}(2022){Quintero Noda}, {Schlichenmaier},
  {Bellot Rubio}, {L{\"o}fdahl}, {Khomenko}, {Jur{\v{c}}{\'a}k}, {Leenaarts},
  {Kuckein}, {Gonz{\'a}lez Manrique}, {Gun{\'a}r}, {Nelson}, {de la Cruz
  Rodr{\'\i}guez}, {Tziotziou}, {Tsiropoula}, {Aulanier}, {Aboudarham},
  {Allegri}, {Alsina Ballester}, {Amans}, {Asensio Ramos}, {Bail{\'e}n},
  {Balaguer}, {Baldini}, {Balthasar}, {Barata}, {Barczynski}, {Barreto
  Cabrera}, {Baur}, {B{\'e}chet}, {Beck}, {Bel{\'\i}o-As{\'\i}n},
  {Bello-Gonz{\'a}lez}, {Belluzzi}, {Bentley}, {Berdyugina}, {Berghmans},
  {Berlicki}, {Berrilli}, {Berkefeld}, {Bettonvil}, {Bianda}, {Bienes
  P{\'e}rez}, {Bonaque-Gonz{\'a}lez}, {Braj{\v{s}}a}, {Bommier}, {Bourdin},
  {Burgos Mart{\'\i}n}, {Calchetti}, {Calcines}, {Calvo Tovar}, {Campbell},
  {Carballo-Mart{\'\i}n}, {Carbone}, {Carlin}, {Carlsson}, {Castro L{\'o}pez},
  {Cavaller}, {Cavallini}, {Cauzzi}, {Cecconi}, {Chulani}, {Cirami},
  {Consolini}, {Coretti}, {Cosentino}, {C{\'o}zar-Castellano}, {Dalmasse},
  {Danilovic}, {De Juan Ovelar}, {Del Moro}, {del Pino Alem{\'a}n}, {del Toro
  Iniesta}, {Denker}, {Dhara}, {Di Marcantonio}, {D{\'\i}az Baso}, {Diercke},
  {Dineva}, {D{\'\i}az-Garc{\'\i}a}, {Doerr}, {Doyle}, {Erdelyi}, {Ermolli},
  {Escobar Rodr{\'\i}guez}, {Esteban Pozuelo}, {Faurobert}, {Felipe}, {Feller},
  {Feijoo Amoedo}, {Femen{\'\i}a Castell{\'a}}, {Fernandes}, {Ferro
  Rodr{\'\i}guez}, {Figueroa}, {Fletcher}, {Franco Ordovas}, {Gafeira},
  {Gardenghi}, {Gelly}, {Giorgi}, {Gisler}, {Giovannelli}, {Gonz{\'a}lez},
  {Gonz{\'a}lez}, {Gonz{\'a}lez-Cava}, {Gonz{\'a}lez Garc{\'\i}a},
  {G{\"o}m{\"o}ry}, {Gracia}, {Grauf}, {Greco}, {Grivel}, {Guerreiro},
  {Guglielmino}, {Hammerschlag}, {Hanslmeier}, {Hansteen}, {Heinzel},
  {Hern{\'a}ndez-Delgado}, {Hern{\'a}ndez Su{\'a}rez}, {Hidalgo}, {Hill},
  {Hizberger}, {Hofmeister}, {J{\"a}gers}, {Janett}, {Jarolim}, {Jess},
  {Jim{\'e}nez Mej{\'\i}as}, {Jolissaint}, {Kamlah}, {Kapit{\'a}n},
  {Ka{\v{s}}parov{\'a}}, {Keller}, {Kentischer}, {Kiselman}, {Kleint},
  {Klvana}, {Kontogiannis}, {Krishnappa}, {Ku{\v{c}}era}, {Labrosse}, {Lagg},
  {Landi Degl'Innocenti}, {Langlois}, {Lafon}, {Laforgue}, {Le Men}, {Lepori},
  {Lepreti}, {Lindberg}, {Lilje}, {L{\'o}pez Ariste}, {L{\'o}pez
  Fern{\'a}ndez}, {L{\'o}pez Jim{\'e}nez}, {L{\'o}pez L{\'o}pez}, {Manso
  Sainz}, {Marassi}, {Marco de la Rosa}, {Marino}, {Marrero}, {Mart{\'\i}n},
  {Mart{\'\i}n G{\'a}lvez}, {Mart{\'\i}n Hernando}, {Masciadri}, {Mart{\'\i}nez
  Gonz{\'a}lez}, {Matta-G{\'o}mez}, {Mato}, {Mathioudakis}, {Matthews}, {Mein},
  {Merlos Garc{\'\i}a}, {Moity}, {Montilla}, {Molinaro}, {Molodij}, {Montoya},
  {Munari}, {Murabito}, {N{\'u}{\~n}ez Cagigal}, {Oliviero}, {Orozco
  Su{\'a}rez}, {Ortiz}, {Padilla-Hern{\'a}ndez}, {Pa{\'e}z Ma{\~n}{\'a}},
  {Paletou}, {Pancorbo}, {Pastor Ca{\~n}edo}, {Pastor Yabar}, {Peat},
  {Pedichini}, {Peixinho}, {Pe{\~n}ate}, {P{\'e}rez de Taoro}, {Peter},
  {Petrovay}, {Piazzesi}, {Pietropaolo}, {Pleier}, {Poedts}, {P{\"o}tzi},
  {Podladchikova}, {Prieto}, {Quintero Nehrkorn}, {Ramelli}, {Ramos Sapena},
  {Rasilla}, {Reardon}, {Rebolo}, {Regalado Olivares}, {Reyes
  Garc{\'\i}a-Talavera}, {Riethm{\"u}ller}, {Rimmele}, {Rodr{\'\i}guez
  Delgado}, {Rodr{\'\i}guez Gonz{\'a}lez}, {Rodr{\'\i}guez-Losada},
  {Rodr{\'\i}guez Ramos}, {Romano}, {Roth}, {Rouppe van der Voort}, {Rudawy},
  {Ruiz de Galarreta}, {Ryb{\'a}k}, {Salvade}, {S{\'a}nchez-Capuchino},
  {S{\'a}nchez Rodr{\'\i}guez}, {Sangiorgi}, {Say{\`e}de}, {Scharmer},
  {Scheiffelen}, {Schmidt}, {Schmieder}, {Scir{\`e}}, {Scuderi}, {Siegel},
  {Sigwarth}, {Sim{\~o}es}, {Snik}, {Sliepen}, {Sobotka}, {Socas-Navarro},
  {Sola La Serna}, {Solanki}, {Soler Trujillo}, {Soltau}, {Sordini}, {Sosa
  M{\'e}ndez}, {Stangalini}, {Steiner}, {Stenflo}, {{\v{S}}t{\v{e}}p{\'a}n},
  {Strassmeier}, {Sudar}, {Suematsu}, {S{\"u}tterlin}, {Tallon}, {Temmer},
  {Tenegi}, {Tritschler}, {Trujillo Bueno}, {Turchi}, {Utz}, {van Harten}, {van
  Noort}, {van Werkhoven}, {Vansintjan}, {Vaz Cedillo}, {Vega Reyes}, {Verma},
  {Veronig}, {Viavattene}, {Vitas}, {V{\"o}gler}, {von der L{\"u}he},
  {Volkmer}, {Waldmann}, {Walton}, {Wisniewska}, {Zeman}, {Zeuner}, {Zhang},
  {Zuccarello}, \& {Collados}}]{EST}
{Quintero Noda}, C., {Schlichenmaier}, R., {Bellot Rubio}, L.~R., {et~al.}
  2022, \aap, 666, A21

\bibitem[{{Rast} {et~al.}(2021){Rast}, {Bello Gonz{\'a}lez}, {Bellot Rubio},
  {Cao}, {Cauzzi}, {Deluca}, {de Pontieu}, {Fletcher}, {Gibson}, {Judge},
  {Katsukawa}, {Kazachenko}, {Khomenko}, {Landi}, {Mart{\'\i}nez Pillet},
  {Petrie}, {Qiu}, {Rachmeler}, {Rempel}, {Schmidt}, {Scullion}, {Sun},
  {Welsch}, {Andretta}, {Antolin}, {Ayres}, {Balasubramaniam}, {Ballai},
  {Berger}, {Bradshaw}, {Campbell}, {Carlsson}, {Casini}, {Centeno}, {Cranmer},
  {Criscuoli}, {Deforest}, {Deng}, {Erd{\'e}lyi}, {Fedun}, {Fischer},
  {Gonz{\'a}lez Manrique}, {Hahn}, {Harra}, {Henriques}, {Hurlburt}, {Jaeggli},
  {Jafarzadeh}, {Jain}, {Jefferies}, {Keys}, {Kowalski}, {Kuckein}, {Kuhn},
  {Kuridze}, {Liu}, {Liu}, {Longcope}, {Mathioudakis}, {McAteer}, {McIntosh},
  {McKenzie}, {Miralles}, {Morton}, {Muglach}, {Nelson}, {Panesar}, {Parenti},
  {Parnell}, {Poduval}, {Reardon}, {Reep}, {Schad}, {Schmit}, {Sharma},
  {Socas-Navarro}, {Srivastava}, {Sterling}, {Suematsu}, {Tarr}, {Tiwari},
  {Tritschler}, {Verth}, {Vourlidas}, {Wang}, {Wang}, {NSO and DKIST Project},
  {DKIST Instrument Scientists}, {DKIST Science Working Group}, \& {DKIST
  Critical Science Plan Community}}]{Rast2021}
{Rast}, M.~P., {Bello Gonz{\'a}lez}, N., {Bellot Rubio}, L., {et~al.} 2021,
  \solphys, 296, 70

\bibitem[{{Reiners} {et~al.}(2016){Reiners}, {Mrotzek}, {Lemke}, {Hinrichs}, \&
  {Reinsch}}]{Reiners2016}
{Reiners}, A., {Mrotzek}, N., {Lemke}, U., {Hinrichs}, J., \& {Reinsch}, K.
  2016, \aap, 587, A65

\bibitem[{{Rimmele} {et~al.}(2020){Rimmele}, {Warner}, {Keil}, {Goode},
  {Kn{\"o}lker}, {Kuhn}, {Rosner}, {McMullin}, {Casini}, {Lin}, {W{\"o}ger},
  {von der L{\"u}he}, {Tritschler}, {Davey}, {de Wijn}, {Elmore}, {Fehlmann},
  {Harrington}, {Jaeggli}, {Rast}, {Schad}, {Schmidt}, {Mathioudakis},
  {Mickey}, {Anan}, {Beck}, {Marshall}, {Jeffers}, {Oschmann}, {Beard},
  {Berst}, {Cowan}, {Craig}, {Cross}, {Cummings}, {Donnelly}, {de Vanssay},
  {Eigenbrot}, {Ferayorni}, {Foster}, {Galapon}, {Gedrites}, {Gonzales},
  {Goodrich}, {Gregory}, {Guzman}, {Guzzo}, {Hegwer}, {Hubbard}, {Hubbard},
  {Johansson}, {Johnson}, {Liang}, {Liang}, {McQuillen}, {Mayer}, {Newman},
  {Onodera}, {Phelps}, {Puentes}, {Richards}, {Rimmele}, {Sekulic}, {Shimko},
  {Simison}, {Smith}, {Starman}, {Sueoka}, {Summers}, {Szabo}, {Szabo},
  {Wampler}, {Williams}, \& {White}}]{DKIST}
{Rimmele}, T.~R., {Warner}, M., {Keil}, S.~L., {et~al.} 2020, \solphys, 295,
  172

\bibitem[{{Robustini} {et~al.}(2019){Robustini}, {Esteban Pozuelo},
  {Leenaarts}, \& {de la Cruz Rodr{\'\i}guez}}]{Robustini2019}
{Robustini}, C., {Esteban Pozuelo}, S., {Leenaarts}, J., \& {de la Cruz
  Rodr{\'\i}guez}, J. 2019, \aap, 621, A1

\bibitem[{{Romano} {et~al.}(2023){Romano}, {Schillir{\`o}}, \&
  {Falco}}]{Romano2023}
{Romano}, P., {Schillir{\`o}}, F., \& {Falco}, M.~i. 2023, arXiv e-prints,
  arXiv:2309.11186

\bibitem[{{Rueedi} {et~al.}(1992){Rueedi}, {Solanki}, \& {Rabin}}]{Ruedi1992}
{Rueedi}, I., {Solanki}, S.~K., \& {Rabin}, D. 1992, \aap, 261, L21

\bibitem[{{Ruiz Cobo} \& {del Toro Iniesta}(1992)}]{RuizCobo1992}
{Ruiz Cobo}, B. \& {del Toro Iniesta}, J.~C. 1992, \apj, 398, 375

\bibitem[{{Sainz Dalda} {et~al.}(2019){Sainz Dalda}, {de la Cruz
  Rodr{\'\i}guez}, {De Pontieu}, \& {Go{\v{s}}i{\'c}}}]{sainz_dalda2019}
{Sainz Dalda}, A., {de la Cruz Rodr{\'\i}guez}, J., {De Pontieu}, B., \&
  {Go{\v{s}}i{\'c}}, M. 2019, \apjl, 875, L18

\bibitem[{{S{\'a}nchez Cuberes} {et~al.}(2005){S{\'a}nchez Cuberes},
  {Puschmann}, \& {Wiehr}}]{SanchezCuberes2005}
{S{\'a}nchez Cuberes}, M., {Puschmann}, K.~G., \& {Wiehr}, E. 2005, \aap, 440,
  345

\bibitem[{{Schad} {et~al.}(2011){Schad}, {Jaeggli}, {Lin}, \&
  {Penn}}]{Schad2011}
{Schad}, T.~A., {Jaeggli}, S.~A., {Lin}, H., \& {Penn}, M.~J. 2011, in
  Astronomical Society of the Pacific Conference Series, Vol. 437, Solar
  Polarization 6, ed. J.~R. {Kuhn}, D.~M. {Harrington}, H.~{Lin}, S.~V.
  {Berdyugina}, J.~{Trujillo-Bueno}, S.~L. {Keil}, \& T.~{Rimmele}, 483

\bibitem[{{Schad} {et~al.}(2015){Schad}, {Penn}, {Lin}, \&
  {Tritschler}}]{Schad2015}
{Schad}, T.~A., {Penn}, M.~J., {Lin}, H., \& {Tritschler}, A. 2015, \solphys,
  290, 1607

\bibitem[{{Schleicher}(1976)}]{T93-27}
{Schleicher}, H. 1976, PhD thesis, University of G\"ottingen

\bibitem[{{Schmidt} {et~al.}(2000){Schmidt}, {Muglach}, \&
  {Kn{\"o}lker}}]{Schmidt2000}
{Schmidt}, W., {Muglach}, K., \& {Kn{\"o}lker}, M. 2000, \apj, 544, 567

\bibitem[{Schmidt {et~al.}(2012)Schmidt, von~der Lühe, Volkmer, Denker,
  Solanki, Balthasar, Gonzalez, Berkefeld, Collados, Hofmann, Kneer, Lagg,
  Puschmann, Schmidt, Sobotka, Soltau, \& Strassmeier}]{schmidt2012}
Schmidt, W., von~der Lühe, O., Volkmer, R., {et~al.} 2012, The GREGOR solar
  telescope on Tenerife

\bibitem[{{Socas-Navarro} {et~al.}(2015){Socas-Navarro}, {de la Cruz
  Rodr{\'\i}guez}, {Asensio Ramos}, {Trujillo Bueno}, \& {Ruiz
  Cobo}}]{Socas2015}
{Socas-Navarro}, H., {de la Cruz Rodr{\'\i}guez}, J., {Asensio Ramos}, A.,
  {Trujillo Bueno}, J., \& {Ruiz Cobo}, B. 2015, \aap, 577, A7

\bibitem[{{Solanki}(2003)}]{Solanki2003}
{Solanki}, S.~K. 2003, \aapr, 11, 153

\bibitem[{{Solanki} {et~al.}(2003){Solanki}, {Lagg}, {Woch}, {Krupp}, \&
  {Collados}}]{Solanki2003b}
{Solanki}, S.~K., {Lagg}, A., {Woch}, J., {Krupp}, N., \& {Collados}, M. 2003,
  \nat, 425, 692

\bibitem[{{Sowmya} {et~al.}(2022){Sowmya}, {Lagg}, {Solanki}, \& {Castellanos
  Dur{\'a}n}}]{Sowmya2022}
{Sowmya}, K., {Lagg}, A., {Solanki}, S.~K., \& {Castellanos Dur{\'a}n}, J.~S.
  2022, \aap, 661, A122

\bibitem[{{Tiwari} {et~al.}(2015){Tiwari}, {van Noort}, {Solanki}, \&
  {Lagg}}]{Tiwari2015}
{Tiwari}, S.~K., {van Noort}, M., {Solanki}, S.~K., \& {Lagg}, A. 2015, \aap,
  583, A119

\bibitem[{{van Noort} {et~al.}(2013){van Noort}, {Lagg}, {Tiwari}, \&
  {Solanki}}]{Tiwari2013}
{van Noort}, M., {Lagg}, A., {Tiwari}, S.~K., \& {Solanki}, S.~K. 2013, \aap,
  557, A24

\bibitem[{{Verma} {et~al.}(2012){Verma}, {Balthasar}, {Deng}, {Liu}, {Shimizu},
  {Wang}, \& {Denker}}]{Verma2012}
{Verma}, M., {Balthasar}, H., {Deng}, N., {et~al.} 2012, \aap, 538, A109

\bibitem[{{Viticchi{\'e}} \& {S{\'a}nchez Almeida}(2011)}]{Viticchie2011}
{Viticchi{\'e}}, B. \& {S{\'a}nchez Almeida}, J. 2011, \aap, 530, A14

\bibitem[{{Westendorp Plaza} {et~al.}(2001){Westendorp Plaza}, {del Toro
  Iniesta}, {Ruiz Cobo}, \& {Mart{\'\i}nez Pillet}}]{Westendorp2001b}
{Westendorp Plaza}, C., {del Toro Iniesta}, J.~C., {Ruiz Cobo}, B., \&
  {Mart{\'\i}nez Pillet}, V. 2001, \apj, 547, 1148

\bibitem[{{Westendorp Plaza} {et~al.}(1998){Westendorp Plaza}, {del Toro
  Iniesta}, {Ruiz Cobo}, {Mart{\'\i}nez Pillet}, {Lites}, \&
  {Skumanich}}]{Westendorp1998}
{Westendorp Plaza}, C., {del Toro Iniesta}, J.~C., {Ruiz Cobo}, B., {et~al.}
  1998, \apj, 494, 453

\bibitem[{{Xu} {et~al.}(2012){Xu}, {Lagg}, {Solanki}, \& {Liu}}]{Xu2012}
{Xu}, Z., {Lagg}, A., {Solanki}, S., \& {Liu}, Y. 2012, \apj, 749, 138

\bibitem[{{Yadav} {et~al.}(2019){Yadav}, {de la Cruz Rodr{\'\i}guez},
  {D{\'\i}az Baso}, {Prasad}, {Libbrecht}, {Robustini}, \& {Asensio
  Ramos}}]{Yadav2019}
{Yadav}, R., {de la Cruz Rodr{\'\i}guez}, J., {D{\'\i}az Baso}, C.~J., {et~al.}
  2019, \aap, 632, A112

\end{thebibliography}

\end{document}